\documentclass[letterpaper]{JHEP3} 

 \usepackage[T1]{fontenc}
 \usepackage[latin9]{inputenc}
\usepackage{color}

\usepackage{array}

\usepackage{setspace}

 \makeatletter

\usepackage{epsfig,multicol,bbm}

\usepackage{amsmath, amssymb, graphics}

\newcommand {\mathsym}[1]{{}}

\usepackage{mathrsfs}
\usepackage{amsfonts}
\usepackage{amsmath}
\usepackage{amssymb}
\usepackage{feynmf}
\usepackage{simplewick}
\usepackage{graphicx}
\usepackage{float}
\usepackage{fancyhdr}
\usepackage{slashed}

\def\be{\begin{equation}}
\def\ee{\end{equation}}
\def\bea{\begin{eqnarray}}
\def\eea{\end{eqnarray}}

\def\nn{\nonumber}
\def\cal{\mathcal}
\def\ph{\phantom}

\global \long \def \f{\phi}
 
\global \long \def \fh{\check{\f}}

\global \long \def \p{\psi}
 
\global \long \def \pt{\tilde{\p}}

\global \long \def \pb{\bar{\psi}}
 
\global \long \def \ptb{\bar{\tilde{\psi}}}

\global \long \def \l{\lambda}
 
\global \long \def \lh{\check{\lambda}}

\global \long \def \Qb{\bar{Q}}

\global \long \def \QQ{\mathcal{Q}}

\global \long \def \Im{\mathcal{I}}
\global \long \def \Jm{\mathcal{J}}

\newcommand\fverb{\setbox\fverbbox=\hbox\bgroup\verb}
\newcommand\fverbdo{\egroup\medskip\noindent%
			\fbox{\unhbox\fverbbox}\ }
\newcommand\fverbit{\egroup\item[\fbox{\unhbox\fverbbox}]}
\newbox\fverbbox

\def\bea{\begin{eqnarray}}
\def\eea{\end{eqnarray}}
\def\be{\begin{equation}}
\def\ee{\end{equation}}

\renewcommand{\[}{\begin{equation}}
\renewcommand{\]}{\end{equation}}

\makeatother

\global\long\global\long\def\m{\mu}

\global\long\global\long\def\f{\phi}

\global\long\global\long\def\e{\epsilon}

\global\long\global\long\def\ad{\dot{\alpha}}

\global\long\global\long\def\bd{\dot{\beta}}

\global\long\global\long\def\la{\lambda}

\global\long\global\long\def\p{\partial}

\global\long\global\long\def\II{\mathcal{I}}

\global\long\global\long\def\JJ{\mathcal{J}}

\global\long\global\long\def\NN{\mathcal{N}}

\global\long\global\long\def\SS{\mathcal{S}}

\global\long\global\long\def\topp#1{\check{#1}}

\global\long\def\fh{\topp{\f}}

\global\long\def\QQ{\mathcal{Q}}

\global\long\def\AA{\mathcal{A}}

\global\long\def\BB{\mathcal{B}}

\global\long\def\CC{\mathcal{C}}

\global\long\def\DD{\mathcal{D}}

\global\long\def\EE{\mathcal{E}}

\global \long \def \a{\alpha}
\global \long \def \b{\beta}
\global \long \def \g{\gamma}
\global \long \def \dt{\delta}
\global \long \def \ad{\dot{\alpha}}
\global \long \def \bd{\dot{\beta}}
\global \long \def \gd{\dot{\gamma}}
\global \long \def \dd{\dot{\delta}}
\global \long \def \ve{\varepsilon}
\global \long \def \e{\epsilon}

\global \long \def \ab{\textbf{a}}
\global \long \def \bb{\textbf{b}}
\global \long \def \cb{\textbf{c}}
\global \long \def \db{\textbf{d}}
\global \long \def \dag{\dagger}
\global \long \def \lbr{\lbrack}
\global \long \def \rbr{\rbrack}
\global \long \def \ph{\phantom}
\global \long \def \pr{\prime}
\global \long \def \dag{\dagger}
\global \long \def \lra{\leftrightarrow}

\title{\begin{center}
The Complete One-Loop Dilation Operator 
 \\ of $\Nm=2$ SuperConformal QCD
\end{center}}

\preprint{YITP-SB-11-16
\\
HU-Mathematik: 2011-9
\\
HU-EP 11/23
}

\author{Pedro Liendo$^{a}$\footnote{Email: pedro.liendo@stonybrook.edu}\;,
$\,$ Elli Pomoni$^{b}$\footnote{Email: pomoni@mathematik.hu-berlin.de} 
$\,$   and   Leonardo Rastelli$^{a}$\footnote{Email: leonardo.rastelli@stonybrook.edu}
\\
\\
\it $^a$ C.N. Yang Institute for Theoretical Physics,\\
\it Stony Brook University, \\
\it Stony Brook, NY 11794-3840, USA
\\
\\
\it $^b$ Institut f\"ur Mathematik und Institut f\"ur Physik, 
\\
Humboldt-Universit\"at zu Berlin\\
Johann von Neumann-Haus, Rudower Chaussee 25, 12489 Berlin, Germany

}
\bigskip

 \abstract{
 \vspace{0.2cm}
 
 We evaluate the full planar one-loop dilation operator of ${\cal N}=2$ SuperConformal QCD,
the $SU(N_c)$ super Yang-Mills theory with $N_f = 2 N_c$ fundamental hypermultiplets, in the flavor-singlet sector.
Remarkably,  the spin-chain Hamiltonian turns out to be completely fixed by superconformal symmetry, as in ${\cal N}=4$ SYM.
We   present a more general calculation, for  the 
superconformal
quiver theory with $SU(N_c) \times SU(N_c)$ gauge group, 
which interpolates between   ${\cal N}=2$ SCQCD 
and the $\mathbb{Z}_2$ orbifold of ${\cal N}=4$ SYM; here
symmetry fixes the Hamiltonian up to a single parameter, corresponding to the ratio of the two marginal gauge couplings.
}
 

\global \long \def \f{\phi} 
\global \long \def \fh{\check{\f}}
\global \long \def \p{\psi} 
\global \long \def \pt{\tilde{\p}}
\global \long \def \pb{\bar{\psi}} 
\global \long \def \ptb{\bar{\tilde{\psi}}}
\global \long \def \l{\lambda} 
\global \long \def \lh{\check{\lambda}}
\global \long \def \Qb{\bar{Q}}
\global \long \def \QQ{\mathcal{Q}}

\global \long \def \Im{\mathcal{I}}
\global \long \def \Jm{\mathcal{J}}
\global \long \def \Nm{\mathcal{N}}
\global \long \def \Fm{\mathcal{F}}
\global \long \def \Bm{\mathcal{B}}
\global \long \def \Cm{\mathcal{C}}
\global \long \def \Em{\mathcal{E}}
\global \long \def \Vm{\mathcal{V}}
\global \long \def \Hm{\mathcal{H}}
\global \long \def \Dm{\mathcal{D}}
\global \long \def \Lm{\mathcal{L}}
\global \long \def \Qm{\mathcal{Q}}
\global \long \def \Km{\mathcal{K}}
\global \long \def \Sm{\mathcal{S}}
\global \long \def \Pm{\mathcal{P}}
\global \long \def \Rm{\mathcal{R}}

\global \long \def \pr{\prime}
\global \long \def \dag{\dagger}

\begin{document} 

\maketitle

\section{Introduction}

Perturbative field theory calculations of the dilation operator have played a crucial
role in uncovering the integrability properties of 
 ${\cal N}=4$ super Yang-Mills (SYM) (see {\it e.g.} \cite{Beisert:2003yb,Staudacher:2004tk,Beisert:2005fw,Beisert:2005tm,Beisert:2006ez, Gromov:2009tv} for
 a partial list of references and \cite{Beisert:2010jr} for a recent comprehensive review).
  As the integrability structure is common to the planar field
 theory and the dual string sigma model, one might even imagine an alternative history
 where the AdS/CFT correspondence is discovered following the hints of the
  field theory integrability.
  
 In this paper we present   a calculation of the complete
planar  one-loop dilation operator of a paradigmatic ${\cal N}=2$ superconformal theory,   the $SU(N_c)$ super Yang-Mills theory with $N_f = 2 N_c$ fundamental hypermultiplets,
in the flavor singlet sector. This theory  is perhaps the simplest $4d$ conformal  field theory 
outside the ``universality class'' of ${\cal N}=4$ SYM and is a very interesting case study.
It admits a large $N$ expansion in the Veneziano sense of $N_f \sim N_c \to \infty$
with $\lambda = g_{YM}^2 N_c$ fixed, and a perturbative expansion in the exactly marginal 't Hooft coupling $\lambda$.
 Is the planar theory integrable? 
Does  it have a dual string description? 
Some progress in answering these two questions, which are logically independent,
was described in \cite{Gadde:2009dj, Gadde:2010zi}. In particular  in \cite{Gadde:2010zi} 
the planar one-loop dilation operator in the scalar subsector was obtained, with some tantalizing hints of integrability. 
As explained in \cite{Gadde:2009dj, Gadde:2010zi}, it is illuminating to embed ${\cal N}=2$ superconformal QCD (SCQCD)  into the ${\cal N}=2$ $SU(N_c) \times SU(N_{\check c})$ quiver theory
(with $N_{\check c} \equiv N_{c}$) which has two independent marginal couplings $g_{YM}$ and $\check g_{YM}$.
The quiver theory interpolates between the standard $\mathbb{Z}_2$ orbifold of ${\cal N}=4$ SYM for $\check g_{YM} = g_{YM}$
 and SCQCD for $\check g_{YM} = 0$.  
 With minor extra work, we  can keep  the calculations in this paper more general and derive
  the full one-loop spin chain Hamiltonian for the whole interpolating quiver theory.
 In the closed subsector of scalar chiral fields the Hamiltonian of the quiver theory has been very recently obtained to three loops \cite{Pomoni:2011}.

  The quiver theory is known to be  integrable
 at the orbifold point $\check g_{YM} = g_{YM}$ \cite{Beisert:2005he}, but it is definitely not integrable
  for generic values of the couplings,
 since the two-body magnon S-matrix does not obey the Yang-Baxter equation \cite{Gadde:2010zi}. It is
 still an open question whether integrability is recovered in the (somewhat singular) SCQCD limit $\check g_{YM} \to 0$.
 We expect the evaluation
 of the full one-loop dilation operator presented here to be a crucial step towards answering this question. 
 
  We find that  the full spin-chain Hamiltonian of ${\cal N}=2$ SCQCD 
  is completely fixed by symmetry,
  as is the case for ${\cal N}=4$ SYM. This came to us as a surprise, because representation theory 
  is less restrictive for the ${\cal N}=2$ superconformal algebra.
   Unlike ${\cal N}=4$ SYM, where each site of the spin chain hosts a single
 ultrashort irreducible representation, in our case single-site letters decompose into three distinct irreps,
 and the tensor product of two nearest-neighbor state spaces has a considerably more intricate decomposition.
 Nevertheless, by a non-trivial generalization of Beisert's approach \cite{Beisert:2003jj, Beisert:2004ry}, we find that symmetry is sufficient 
 to determine the Hamiltonian up to overall normalization. We regard this as a hint to
  deeper solvability/integrability properties than meet the eye. The generalization to the interpolating quiver
  is then as simple as one may hope: symmetry leaves a single undetermined parameter, which gets identified
  with the ratio of the two marginal gauge couplings.

After reviewing some basics and setting  notations in Section 2, we describe the strategy of our calculation
in Section 3. Following Beisert \cite{Beisert:2003jj, Beisert:2004ry}, the  evaluation of the full one-loop dilation operator
consists of two parts. First, one computes the dilation operator in a closed  subsector with $SU(1,1)$ symmetry;
then one uses superconformal symmetry to uplift the result to the full theory. The details
are considerably more involved than in the ${\cal N}=4$ case: the two-site state space is spanned by 
a baroque list of  irreducible representations, some of which appear in different copies, leading to an intricate mixing problem. 
Nevertheless, we are able to identify a suitable subsector, 
whose Hamiltonian uplifts to the full theory. The complete Hamiltonian is written as a sum of two-site superconformal projectors.

We compute the Hamiltonian in the closed subsector both by direct evaluation of Feynman diagrams (Section 4)
 and by a purely algebraic approach using the constraints of superconformal symmetry (Section 5).
 The algebraic method is similar to the one used by Beisert in his thesis \cite{ Beisert:2004ry} for ${\cal N}=4$ SYM, and rather surprisingly
 leads to a similar uniqueness result. In both cases the key feature is the existence of a centrally-extended  $SU(1|1)$ symmetry, 
 which commutes with the bosonic $SU(1,1)$ symmetry up to local gauge transformations on the spin chain. 
  Finally in Section 6 we re-write the Hamiltonian, so far expressed  rather implicitly as a sum over superconformal projectors, in the much more explicit 
 ``harmonic action'' \cite{Beisert:2003jj} form,  which is easy to  implement on any given state. Algebraic techniques to obtain spin-chain Hamiltonians were also used in \cite{Beisert:2003ys, Zwiebel:2005er} for $\Nm=4$ SYM at higher loops and in \cite{Zwiebel:2009vb} for the ABJM theory.
 
 The interpolating quiver theory, while not integrable, is interesting in its own right. It has
 a dual string description as the IIB background  $AdS_5 \times S^5/\mathbb{Z}_2$, with a non-trivial period of $B_{NSNS}$ through the collapsed
 cycle of the orbifold \cite{Kachru:1998ys, Klebanov:1999rd}. 
For generic values of the couplings the symmetry of the spin chain in the excitation picture
 contains a single copy of the supergroup $SU(2|2)$ (as opposed to the two independent $SU(2|2)$s of the ${\cal N}=4$ chain). 
 The two-body S-matrix of magnons transforming under the surviving $SU(2|2)$ can be determined to all orders in the gauge coupling \cite{Beisert:2005tm, Gadde:2010ku}, up to an overall
 phase ambiguity, from symmetry considerations alone.
 Armed with the explicit one-loop  Hamiltonian, in Appendix D we confirm the prediction of \cite{Gadde:2010ku} to lowest order in the coupling.
 Three other technical appendices complement the text.

\section{Preliminaries}

We begin by quickly reviewing ${\cal N}=2$ superconformal QCD, the closely related $\mathbb{Z}_2$ quiver theory, 
and the structure of their spin chains. For more details, including the explicit Lagrangians,
we refer to \cite{Gadde:2010zi}.

\subsection{Field Content and Symmetries}

We summarize in Table 1 the field content and quantum numbers
of the ${\cal N}=2$  SYM theory with gauge group $SU(N_{c})$ and $N_{f} = 2 N_c$ fundamental hypermultiplets,
which we refer to as ${\cal N} = 2$ superconformal QCD (SCQCD). 
Its global symmetry group is $U(N_f)  \times SU(2)_R  \times U(1)_r$,
where  $SU(2)_R  \times U(1)_r$  is the R-symmetry subgroup of the superconformal group. 
We use indices $\alpha, \beta = \pm$ and $\dot \alpha, \dot \beta = \dot \pm$ for the Lorentz group,
 $ \II,  \JJ = 1,2$  for $SU(2)_R$, $i, j=1,  \dots N_f$ for the flavor group $U(N_f)$
and $a,b=1,  \dots N_c$ for the color group $SU(N_c)$.
The $ \NN=2$ vector multiplet consists of a gauge field $A_{\mu}$, two Weyl spinors
$ \lambda_{\II \alpha}$, $ \II= 1,2$, which form a doublet under $SU(2)_{R}$,
and one complex scalar $ \phi$, all in the adjoint representation of $SU(N_{c})$. 
Each $ \NN=2$ hypermultiplet consists of 
 an $SU(2)_{R}$ doublet $Q_{\II}$ of complex scalars
and of two Weyl spinors $ \psi_{\alpha}$ and $ \tilde{\psi}_{\alpha}$, $SU(2)_{R}$ singlets.

 \begin{table}
 \begin{centering}
 \begin{tabular}{|c||c|c|c|c|}
 \hline 
 & $SU(N_{c})$  & $U(N_{f})$  & $SU(2)_{R}$  & $U(1)_{r}$ \tabularnewline
 \hline
 \hline 
$\Qm^{\ph{\a}\Im}_{\a}$  &  \textbf{$ \mathbf{1}$}  &  \textbf{$ \mathbf{1}$}  &  \textbf{$ \mathbf{{2}}$}  & $+1/2$ \tabularnewline
 \hline 
$\Sm^{\ph{\Im}\a}_{\Im}$ &  \textbf{$ \mathbf{1}$}  &  \textbf{$ \mathbf{1}$}  &  \textbf{$ \mathbf{2}$}  & $-1/2$ \tabularnewline
 \hline
 \hline 
$A_{\m}$  & Adj  &  \textbf{$ \mathbf{1}$}  &  \textbf{$ \mathbf{1}$}  & $0$ \tabularnewline
 \hline 
$ \f$  & Adj  &  \textbf{$ \mathbf{1}$}  &  \textbf{$ \mathbf{1}$}  & $-1$ \tabularnewline
 \hline 
$ \la_{\II\a}$  & Adj  &  \textbf{$ \mathbf{1}$}  &  \textbf{$ \mathbf{2}$}  & $-1/2$ \tabularnewline
 \hline 
$Q_{\II}$  & $ \Box$  & $ \Box$  &  \textbf{$ \mathbf{2}$}  & $0$ \tabularnewline
 \hline 
$ \psi_{\alpha}$  & $ \Box$  & $ \Box$  &  \textbf{$ \mathbf{1}$}  & $+1/2$ \tabularnewline
 \hline 
$ \tilde{\psi}_{\alpha}$  & $ \overline{\Box}$  & $ \overline{\Box}$  &  \textbf{$ \mathbf{1}$}  & $+1/2$ \tabularnewline
 \hline
 \end{tabular}
 \par \end{centering}
 \caption{Field content and symmetries of  $ \NN=2$ SCQCD.
We show the quantum numbers of  the Poincar\'e supercharges  $\Qm^{\,\,\,\Im}_{\a}$, 
of the conformal supercharges $\Sm^{\,\,\,\a}_{\Im}$ and
of the elementary component fields. 
Conjugate objects (such as $ \tilde {\cal Q}_{\II\,  \dot{\alpha}}  $ and $ \bar  \phi$) are not written explicitly.}
\label{charges}
 \end{table}

${\cal N} = 2$ SCQCD, which has one exactly marginal coupling $g_{YM}$,
can be viewed as a limit of the
 $\NN=2$  $\mathbb{Z}_2$ quiver theory with gauge group\footnote{The gauge groups are identical, $N_{\check c} \equiv N_c$,
 but we find it useful to distinguish with a ``check'' all the quantities pertaining to the second gauge group.} 
 $SU(N_c) \times SU(N_{\check c})$, which has two exactly marginal couplings  $g_{YM}$ and $\check g_{YM}$,
 as  $\check g_{YM} \to 0$.
 When $g_{YM} = \check g_{YM}$ the quiver theory is the familiar $\mathbb{Z}_2$ orbifold of ${\cal N} =4$ SYM.  
Table 2 summarizes the field content and symmetries of the quiver theory.
Besides the R-symmetry group $SU(2)_R \times U(1)_r$, the theory has an additional
 $SU(2)_L$ global symmetry, whose indices we denote by $\hat {\cal I}, \hat {\cal J} = \hat 1, \hat 2$.
Supersymmetry organizes the component fields  into the $\NN =2$ vector multiplets of each factor of the gauge group,
 $(\phi, \lambda_\II, A_\mu)$ and  $(\check \phi, \check \lambda_\II,\check A_\mu)$,
 and into two bifundamental hypermultiplets,  $(Q_{\II, \hat 1}, \psi_{\hat 1}, \tilde \psi_{\hat 1})$
 and  $(Q_{\II, \hat 2}, \psi_{\hat 2}, \tilde \psi_{\hat 2})$. 

\begin{table}
\begin{centering}
\begin{tabular}{|c||c|c|c|c|c|}
\hline 
 & $SU(N_{c})$  & $SU(N_{\check c})$  & $SU(2)_{R}$  & $SU(2)_{L}$  & $U(1)_{R}$\tabularnewline
\hline
\hline 
$\Qm^{\ph{\a}\Im}_{\a}$ & \textbf{${\bf 1}$}  & \textbf{${\bf 1}$}  & \textbf{${\bf 2}$}  & \textbf{${\bf 1}$}  & +1/2\tabularnewline
\hline 
$\SS_{\II}^{\,\,\, \alpha}$ & \textbf{${\bf 1}$}  & \textbf{${\bf 1}$}  & \textbf{${\bf 2}$}  & \textbf{${\bf 1}$}  & --1/2\tabularnewline
\hline
\hline 
$A_{\mu}$  & Adj  & \textbf{${\bf 1}$}  & \textbf{${\bf 1}$}  & \textbf{${\bf 1}$}  & 0\tabularnewline
\hline 
$\topp A_{\mu}$  & \textbf{${\bf 1}$}  & Adj  & \textbf{${\bf 1}$}  & \textbf{${\bf 1}$}  & 0\tabularnewline
\hline 
$\f$  & Adj  & \textbf{${\bf 1}$}  & \textbf{${\bf 1}$}  & \textbf{${\bf 1}$}  & --1\tabularnewline
\hline 
$\topp{\f}$  & \textbf{${\bf 1}$}  & Adj  & \textbf{${\bf 1}$}  & \textbf{${\bf 1}$}  & --1\tabularnewline
\hline 
$\la_{{\II} \alpha }$  & Adj  & \textbf{${\bf 1}$}  & \textbf{${\bf 2}$}  & \textbf{${\bf 1}$}  & --1/2\tabularnewline
\hline 
$\topp{\la}_{\II \a}$  & \textbf{${\bf 1}$}  & Adj  & \textbf{${\bf 2}$}  & \textbf{${\bf 1}$}  & --1/2\tabularnewline
\hline 
$Q_{\II\hat{\II}}$  & $\Box$  & $\overline{\Box}$  & \textbf{${\bf 2}$}  & \textbf{${\bf 2}$}  & 0\tabularnewline
\hline 
$\psi_{\hat{\II} \,\alpha}$  & $\Box$  & $\overline{\Box}$  & \textbf{${\bf 1}$}  & \textbf{${\bf 2}$}  & +1/2\tabularnewline
\hline 
$\tilde{\psi}_{\hat{\II} \,\alpha}$  & $\overline{\Box}$  & $\Box$  & \textbf{${\bf 1}$}  & \textbf{${\bf 2}$}  & +1/2\tabularnewline
\hline
\end{tabular}
\par\end{centering}
\caption{\label{orbifoldcharges}
Field content and symmetries of  the quiver theory that interpolates between the $\mathbb{Z}_2$ orbifold
of ${\cal N}=4$ SYM and ${\cal N}=2$ SCQCD.
}
\end{table}

Setting $\check g_{YM} = 0$, the second vector multiplet $(\check \phi, \check \lambda_\II, \check A_\mu)$ becomes free and completely decouples 
from the rest of the theory, which coincides 
with ${\cal N} = 2$ SCQCD (the field content is the same and ${\cal N} = 2$ susy does the rest). 
The $SU(N_{\check{c}})$
symmetry can now be interpreted as a global flavor symmetry. In fact there is a symmetry enhancement
$SU(N_{\check c}) \times SU(2)_L \to U(N_f = 2 N_c)$: 
 the $SU(N_{\check c})$ index $\check a$
and the $SU(2)_L$ index $\hat \II$
 can  be combined  into a single flavor index $i \equiv (\check a, \hat I) =1, \dots 2 N_c$.

We  work in the large $N_c \equiv N_{\check c}$ limit, keeping fixed the 't Hooft  couplings
\be
\lambda \equiv g_{YM}^2 N_c  \equiv 8 \pi^2 g^2 \, , \qquad  \check \lambda \equiv \check g_{YM}^2 N_{\check c}  \equiv 8 \pi^2 \check g^2 \, .
\ee
We will often refer to the theory with arbitrary $g$ and $\check g$ as the ``interpolating SCFT'', thinking of keeping $g$ fixed
as we vary $\check g$ from  $\check g = g$  (orbifold theory) to  $\check g = 0$ 
 (${\cal N} =2$ SCQCD  $\oplus$ extra  $N_{\check c}^2-1$ free vector multiplets).

\subsection{The Spin Chain}

As familiar, the planar dilation operator of a gauge theory
can be represented as the Hamiltonian of a spin chain.
Each site of the chain is occupied by a ``letter'' of the gauge theory:
a letter ${\cal D}^k {\cal A}$ can be any of the elementary fields ${\cal A}$ acted on by an arbitrary number of gauge-covariant derivatives ${\cal D}$.
A closed chain corresponds to a single-trace operator.

In the interpolating SCFT, letters belonging to the vector multiplets are in the adjoint representation of 
either gauge group (index structures $^a_{\ph{\l\l}b}$ and  $^{\check{a}}_{\ph{\l\l}\check{b}}$), while letters belonging to the hypermultiplets are in a bifundamental representation
(index structures $^a_{\ph{\l\l}\check{b}}$ and  $^{\check{a}}_{\ph{\l\l}b}$).  In SCQCD, vector letters 
have index structure  $^a_{\ph{\l\l}b}$, while hyper letters have stuctures  $^a_{\ph{\l\l}i}$ and $^i_{\ph{\l\l}b}$. We  restrict attention 
to the flavor-singlet sector of SCQCD. Then, as explained in \cite{Gadde:2009dj,Gadde:2010zi}, in the Veneziano limit of $N_c \to \infty$, $N_f \to \infty$ with $N_f/N_c \equiv 2$
and $g_{YM}^2 N_c$ fixed,  the basic building blocks are the  ``generalized single-trace operators'', where consecutive
letters have contracted color or flavor indices, for example
\be
\label{local-operator}
\mbox{Tr}[\bar{\phi}\phi\phi Q_{\II}\bar{Q}^{\JJ}\bar{\phi}] =    \bar \phi^a_{\, \, \,b} \phi^b_{\, \, \,c}  \phi^c_{\, \, \,d}  Q^{\, d}_{\II \, \, i}  \bar Q^{\JJ  i}_{\quad e} \bar \phi^e_{\, \, \, a}    \, , \quad a,b,c,d,e=1,\dots N_c\, ,\quad i = 1, \dots N_f \,.
\ee
In the large $N$ Veneziano limit the action of the dilation operator is well-defined on
generalized single-traces, because mixing with multi-traces is suppressed.
We write the planar dilation operator as
\be
D = g^2 H \, ,
\ee
where $H$ is the spin-chain  Hamiltonian. 
At one-loop, $H$ is of nearest-neighbor form,
\be
H = \sum_{\ell=1}^L H_{\ell ,\ell+1} \,.
\ee
The one-loop Hamiltonian of the interpolating theory depends on the ratio of the couplings, $\kappa \equiv \check g/g$,
while the one-loop Hamiltonian of SCQCD has no parameters. We can obtain $H_{SCQCD}$ as the $\kappa \to 0$
limit of the interpolating Hamiltonian, restricted to the $U(N_f)$ singlet subsector (consecutive $SU(2)_L$ indices are contracted).
 
\section{Lifting the Full One-loop Hamiltonian from a Subsector}
\label{liftingS}

 Computing 
the complete one-loop Hamiltonian appears to be a daunting combinatorial task, because of the sheer number of possible
two-letter structures on which the Hamiltonian can act.  
For ${\cal N}=4$ SYM, Beisert \cite{Beisert:2003jj} was able to determine the full one-loop Hamiltonian by making maximal use
of the power of superconformal symmetry. 
The letters of  ${\cal N}=4$ SYM belong to a single representation
of the superconformal algebra, the  ultrashort ``singleton'' representation $V_F$. 
The tensor product of two singletons  has a simple decomposition into an infinite sum of irreducible representations, 
\be \label{doubleton}
V_F \times V_F= \sum_{j=0}^\infty V_j \, .
\ee
The one-loop Hamiltonian can then be written as 
\be
H_{1 2} = \sum_{j=0}^\infty  f(j) \, {\cal P}_{j}\, ,
\ee
where  $P_{j}$ is a projector on the $V_j$ module for letters at sites 1 and 2.
Beisert's strategy was to identify a simple closed subsector of the theory,
such that each of the $V_j$ modules contains a representative within the subsector.
The coefficients $f(j)$ and thus the full Hamiltonian can be read off from the Hamiltonian
of the closed subsector. A particularly clever choice \cite{Beisert:2004ry} of subsector is the $SU(1,1) \times U(1|1)$ subsector 
comprising the letters $ D_{+ \dot +}^n \lambda_{+ }$, where $\lambda_\alpha$ is one of the four Weyl fermions.
The algebraic constraints of superconformal symmetry are so powerful that they fix
the Hamiltonian of this sector, up to the overall normalization which corresponds to a rescaling of the coupling.\footnote{In his first  calculation \cite{Beisert:2003jj}, Beisert considered 
the $SU(1,1)$ subsector consisting of the letters ${\cal D}_{+ \dot +}^n Z$, where $Z$ is a complex scalar,
and determined the $SU(1,1)$ one-loop Hamiltonian  by direct evaluation of Feynman diagrams.} 
All in all, the one-loop Hamiltonian of ${\cal N}=4$ SYM is determined by superconformal symmetry alone.

In adapting Beisert's strategy to our case, we are faced with  the complication that
the letters belong to three distinct representations of the ${\cal N}=2$ superconformal algebra,
with their tensor products containing  different copies of the same module. This leads to a rather
intricate mixing problem. Nevertheless, the problem turns out to be tractable.
We are able to identify a subsector from which the full Hamiltonian can be lifted.
We have determined the Hamiltonian within the subsector both by explicit Feynman
diagram calculations, as described in Section \ref{perturbativeS}, and by exploiting the constraints of the superconformal algebra,
as described in Section \ref{algebraicS}.

\subsection{Superconformal Projectors}

Our notations for superconformal representations
are borrowed from \cite{Dolan:2002zh} and reviewed in Appendix \ref{multipletsA}. The letters of SCQCD (as well as of the whole interpolating theory)
belong to three superconformal representations, which we denote by ${\cal H}$, ${\cal V}$ and $\bar {\cal V}$.
The hypermultiplet letters ($Q_{\cal I}$ and its descendants\footnote{We are suppressing for now  $SU(2)_L$ indices, since $SU(2)_L$
commutes with the superconformal algebra.}) 
belong to the representation ${\cal H} \equiv \hat {\cal B}_{\frac{1}{2}}$, while the vector multiplet letters
split into the two conjugate representations ${\cal V}\equiv \bar {\cal E}_{1 (0,0)}$ ($\phi$ and its descendants) and
 $\bar {\cal V}\equiv {\cal E}_{1 (0,0)}$ ($\bar \phi$ and its descendants). It is not difficult, using ${\cal N}=2$ superconformal characters\footnote{See for example
 \cite{Bianchi:2006ti} for an illustration 
 of superconformal character techniques in ${\cal N}=4$ case.}, to evaluate the relevant tensor products\footnote{Following \cite{Dolan:2002zh}, we extend
 the definition of the $\hat {\Cm}$ multiplets to $j_1,j_2=-\frac{1}{2}$ according to the rules: \\$\hat{\Cm}_{0(-\frac{1}{2},-\frac{1}{2})} \equiv \hat{\Bm}_1$,    
 $\hat{\Cm}_{0(0,-\frac{1}{2})}\equiv \bar{\Dm}_{\frac{1}{2}(0,0)}$,   
 $\hat{\Cm}_{0(-\frac{1}{2},0)}\equiv{\Dm}_{\frac{1}{2}(0,0)}$,  $\hat{\Cm}_{0(\frac{1}{2},-\frac{1}{2})}\equiv\bar{\Dm}_{\frac{1}{2}(\frac{1}{2},0)}$ and $\hat{\Cm}_{0(-\frac{1}{2}, \frac{1}{2})}\equiv\Dm_{\frac{1}{2}(0,\frac{1}{2})}$. }
\bea
\label{HxH}
\Hm \times \Hm & = & \sum_{q=-1}^{\infty} \hat{\Cm}_{0(\frac{q}{2},\frac{q}{2})}\, ,
\\
\label{HxV}
\Hm \times \Vm & = & \sum_{q=-1}^{\infty}\hat{\Cm}_{0(\frac{q+1}{2},\frac{q}{2})}= \Vm \times  \Hm \, ,
\\
\label{HxVb}
\Hm \times \bar{\Vm} & = & \sum_{q=-1}^{\infty}\hat{\Cm}_{0(\frac{q}{2},\frac{q+1}{2})} = \bar \Vm \times  \Hm \, ,
\eea
\bea
\Vm \times \Vm  & = & \bar{\Em}_{2(0,0)} 
+ \sum_{q=0}^{\infty}\hat{\Cm}_{0(\frac{q+1}{2},\frac{q-1}{2})}\, ,
\\
\bar{\Vm} \times \bar{\Vm} & = &  \Em_{2(0,0)}
+ \sum_{q=0}^{\infty}\hat{\Cm}_{0(\frac{q-1}{2},\frac{q+1}{2})}\, ,
\\
\label{VxVb}
\Vm \times \bar{\Vm} & = &   \sum_{q=0}^{\infty}\hat{\Cm}_{0(\frac{q}{2},\frac{q}{2})}= \bar \Vm \times {\Vm} \,.
\eea
The two-site Hamiltonian $H_{12}$ can still be written as a sum of superconformal projectors,
but we must  take into account mixing between different sectors. For example, since the representation
$ \hat{\Cm}_{0(\frac{q}{2},\frac{q}{2})}$ appears in the tensor products  $\Hm \times \Hm$, $\Vm \times \bar{\Vm}$ and $\bar \Vm \times {\Vm}$,
these states will mix. The restriction of $H_{12}$ to this subspace  takes the form
\be 
H_{12} = A_{11}(-1) \, {\cal P}_{(-\frac{1}{2}, -\frac{1}{2})} +\sum_{q=0}^{\infty} 
\left(
\begin{array}{ccc}
A_{11}(q) & A_{12}(q) & A_{13}(q)\\
A_{21}(q) & A_{22}(q) &A_{23}(q) \\
A_{31}(q) & A_{32}(q) &A_{33}(q) \\
\end{array}\right){\cal P}_{(\frac{q}{2}, \frac{q}{2})} \, ,
\ee
where for each $q$ the $3 \times 3$ matrix $A_{rs}(q)$ is the mixing matrix of   $\Hm \times \Hm$, $\Vm \times \bar{\Vm}$ and $\bar \Vm \times {\Vm}$.
Similarly, there is mixing between ${\cal H} \times {\cal V}$ and ${\cal V} \times {\cal H}$, and between ${\cal H} \times \bar {\cal V}$ and $\bar {\cal V} \times {\cal H}$, but
 no mixing for either ${\cal V} \times {\cal V}$ and $\bar {\cal V} \times \bar {\cal V}$, since these latter products decompose into representations that do not  appear anywhere else.

\subsection{A Convenient Subsector}

A straightforward way to obtain the coefficients that multiply the superconformal projectors
 would be to evaluate the dilation operator on the superconformal primaries of each module. The projectors act trivially on these states and the mixing matrix could be read immediately. However, the primaries are complicated objects (see Appendix \ref{PrimariesA}) and it will be easier to consider certain descendants instead. 
 
  We have identified a closed subsector, somewhat analogous to the  $SU(1,1) \times U(1|1)$ subsector  \cite{Beisert:2004ry} of ${\cal N}=4$ SYM.
 In SCQCD, our subsector consists of the letters $\lambda_{2 +}$,  $\bar \lambda_{2 \dot +}$,
 $Q_2$ and $\bar Q_2$,  acted upon by an arbitrary number of covariant derivatives ${\cal D}_{+ \dot +}$. Note that all the $SU(2)_R$ indices
 are taken to be subscripts\footnote{If the natural position of the $SU(2)_R$ index is as a superscript,
 as in $\bar{\lambda}_{\ad}^\Im$ and $\bar Q^{\cal I}$, we lower it using  $\epsilon_{{\cal I} {\cal J}}$.} with the value ${\cal  I}=2$.
 In the interpolating theory, we add  $\check \lambda_{2  +}$ and $\bar {\check \lambda}_{2 \dot +}$ to the list.
 It will be convenient to define (with ${\cal D}\equiv {\cal D}_{+ \dot +}$)
 \bea
\label{l}
\lambda_k = \frac{\Dm^k}{k!} \lambda_{ 2 +}\,, &  \quad & \bar{\lambda}_k  = \frac{\Dm^k}{k!}  \bar{\lambda}_{ 2 \dot +}\,,
\\
\label{lc}
\lh_k = \frac{\Dm^k}{k!} \lh_{2 +}\,, & \quad & \bar{\lh}_k = \frac{\Dm^k}{k!}{\bar{\hat \lambda} }_{2\dot +}\,,
\\
\label{Q}
Q_{k\, \hat {\cal I}} = \frac{\Dm^k}{k!} Q_{2\,  \hat {\Im}}\,, &\quad  & \bar{Q}_k^{\hat{\Im}} = \frac{\Dm^k}{k!} \bar{Q}_{\ph{\Im}2}^{\hat {\cal I}}\,.
\eea
The  $SU(2)_L$ indices $\hat {\cal I} = \hat 1, \hat 2$ will often be suppressed to avoid cluttering.

The sector   (\ref{l})-(\ref{Q})   is closed to all loops, as one easily checks by using conservation of  the engineering dimension and of the Lorentz and the  R-symmetry quantum numbers.
The subgroup of the superconformal group acting on the sector is $SU(1,1) \times SU(1|1) \times SU(1|1) \times U(1)$. The 
$SU(1,1)$ generators are
\bea \label{SU(1,1)}
	\Jm'_+(g) & = & \Pm_{+ \dot{+}}(g)\, ,
	\\
	\Jm'_-(g) & = & \Km^{+\dot{+}}(g)\, ,
	\\
	\Jm'_3(g) & = & \frac{1}{2}D_0+\frac{1}{2}\delta D(g) + \frac{1}{2}\Lm^{\ph{+}+}_{+} + \frac{1}
	{2}\dot{\Lm}^{\ph{+}\dot{+}}_{\dot{+}} \, ,
	\eea
where $\delta D(g) \equiv D(g)-D_0$ is the difference between the quantum dilation operator and its classical limit $D_0 = D(0)$.
The states $Q_{k=0}$ and $\bar Q_{k=0}$   are primaries of spin $-\frac{1}{2}$ representations of $SU(1,1)$,
while the states $\lambda_{k=0}$, $\check \lambda_{k=0}$, $\bar \lambda_{k=0}$, $\bar {\check \lambda}_{k=0}$
are primaries of spin $-1$ representations of $SU(1,1)$.
The $SU(1|1)\times SU(1|1) \times U(1)$ generators will be presented in Section \ref{algebraicS}, they play a key role in the algebraic  approach
but will not be important for the analysis of  the next Section.

Each of the modules appearing on the right hand side of the tensor products (\ref{HxH})-(\ref{VxVb}) contains a representative in this subsector. The representatives are primaries of $SU(1,1)$,
and  descendants with respect to the full $SU(2,2|2)$. This is sufficient to uplift   the Hamiltonian of the subsector to the full Hamiltonian.

\section{Field Theory Evaluation of the Hamiltonian}
\label{perturbativeS}

In this section we describe the field-theory evaluation of the one-loop Hamiltonian in the $SU(1,1) \times SU(1|1) \times SU(1|1) \times U(1)$ subsector,
and its uplifting to the full Hamiltonian.
We present the result for the interpolating theory, as a function of $\kappa = \check g /g$.
 The result for SCQCD is obtained by taking the limit
$\kappa \to 0$ and focussing on the relevant subspace (that is, discarding the ``checked'' fields and contracting adjacent  $SU(2)_L$ indices).
We can focus on evaluating the Hamiltonian on two-site states with open indices    $^a_{\ph{\l\l}b}$ and $^a_{\ph{\l\l}\check{b}}$, since
 the Hamiltonian acting on the structures  
$^{\check{a}}_{\ph{\l\l}\check{b}}$ and  $^{\check{a}}_{\ph{\l\l}b}$ 
 is immediately obtained by interchanging $g \leftrightarrow 
\check{g}$.

\subsection{$\Vm \times \Vm$}
\label{VxVFeynman}

\begin{figure}[h]
    \begin{center}
    \includegraphics[scale=0.6]{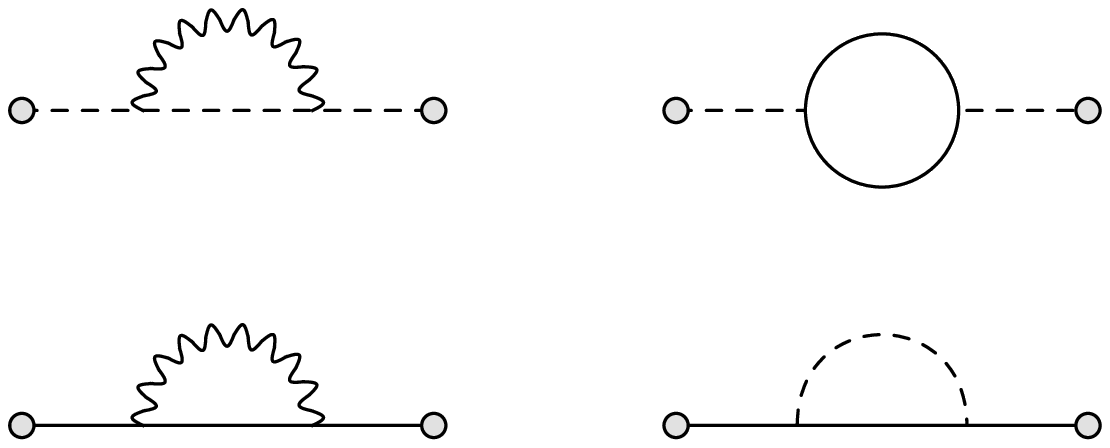}                     
    \caption{Self-energy corrections
    of the external legs.
    Full lines denote fermion propagators,
    curly lines gauge boson propagators and dashed lines scalar propagators. For $\l_k \l_{n-k}$ mixing only the second line of 
    diagrams contributes. Corrections to bosonic legs depicted in the first line will be relevant in the next two subsections.}
    \label{DiagrSelf}
    \end{center}
    \end{figure}

\begin{figure}[h]
    \begin{center}
    \includegraphics[scale=0.6]{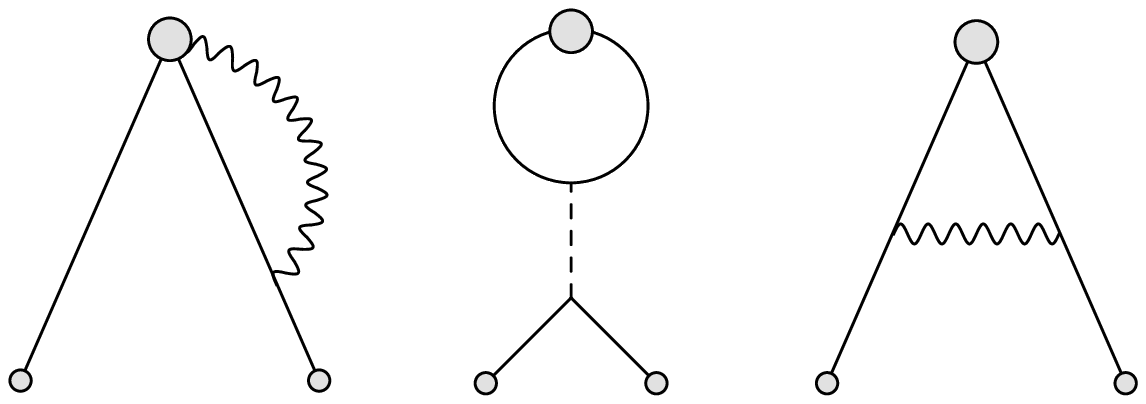}                     
    \caption{One-loop 1PI Feynman diagrams contributing to  $\l_k \l_{n-k}$ mixing. 
     The last two diagrams are in fact zero 
              for this combination of Lorentz and $SU(2)_R$ indices.}
    \label{DiagrVV}
    \end{center}
    \end{figure}

The states of the $SU(1,1) \times SU(1|1) \times SU(1|1) \times U(1)$  subsector belonging to $\Vm \times \Vm$ have
the form $\l_k \l_{n-k}$. The relevant Feynman diagrams are shown in Figures \ref{DiagrSelf} and  \ref{DiagrVV}. All our calculations are done in Feynman gauge where the gauge propagator reads $\frac{g_{\mu \nu}}{k^2}$. A sample field theory calculation is described in Appendix \ref{sampleA}.

The action of the Hamiltonian on these states is
\be 
H'_{12} \l_k \l_{n-k}=2\sum^n_{k'=0}c_{n,k,k^{\pr}}\l_{k^{\pr}}\l_{n-k^{\pr}} \, ,
\ee
with
\be 
c_{n,k,k'}=\delta_{k=k^\pr}\left(h(k+1)+h(n-k+1)\right) -\frac{\delta_{k\neq k^\pr}}{|k-k^\pr|}+\frac{\delta_{k>k'}}{n-k'+1}+\frac{\delta_{k< k'}}{k'+1}\, ,
\ee
where $h(k)$ are the harmonic numbers, $h(k)=\sum_{j=1}^k\frac{1}{j}$ and $h(0)\equiv 0$.

Using the oscillator representation (see Appendix \ref{oscillatorsA}) it is easy to check that $H'_{12}$ is invariant under $SU(1,1)$. 
We can then write the Hamiltonian density as 
\be 
H'_{12}=\sum_{j=0}^{\infty}A (j)\Pm'_{-1-j}\, ,
\ee
where $\Pm'_{-1-j}$ is a projector on  the $SU(1,1)$ module of spin $-1-j$.
To obtain the coefficients $A(j)$ we act on the $SU(1,1)$ highest weights,
\be 
\Jm(j)=-\frac{(j+2)}{(j+1)}\sum_{k=0}^j\frac{(-1)^k}{k+1}\binom{j}{k}\binom{j+1}{k}\Dm^{j-k}\l_{2+} \Dm^k \l_{2+}\, .
\ee
The result is
\be
H'_{12}\Jm(j)=4h(j+1)\Jm(j) \, ,
\ee
which implies  $A(j)=4h(j+1)$. The lifting procedure is now straightforward: $\Jm(j)$ is not only an $SU(1,1)$ highest weight but also a superconformal descendant, it can be obtained by applying $-\frac{1}{2}\Rm_{2}^{\ph{2}1}\Qm_{+}^{\ph{+}2}$ to (\ref{VVprimaryq0}) for $j=0$ 
and $\Qm_{+}^{\ph{+}1}\tilde{\Qm}_{\dot{+}2}$ to (\ref{VVprimaryq}) for $j>0$.
The $SU(1,1)$ modules are sub-modules of the the superconformal modules with $j=q$. The only module not present in this sub-sector is $\bar{\Em}_{2(0,0)}$, but we know that
this is a protected multiplet so its coefficient is just zero. All in all, the Hamiltonian density in ${\cal V} \times {\cal V}$ is
\be
\label{FullHonVxV} 
H_{12}= 0 \times \Pm_{\bar{\Em}}+\sum_{q=0}^{\infty}4h(q+1)\Pm_{(\frac{q+1}{2},\frac{q-1}{2})} \,.
\ee

\subsection{$\Vm \times \Hm \leftrightarrow \Hm \times {\Vm}$}
\label{VxHFeynman}

The mixing between $\Vm \times \Hm \leftrightarrow \Hm \times {\Vm}$ and $\bar{\Vm} \times \Hm \leftrightarrow \Hm \times {\bar{\Vm}}$ should be identical, we only need to focus on the first case. The relevant states  are  $\l_k Q_{n-k} \in \Vm \times \Hm$ and  $Q_{k}\lh_{n-k}   \in \Hm \times {\Vm}$. 
	\begin{figure}[h]
    \begin{center}
    \includegraphics[scale=0.6]{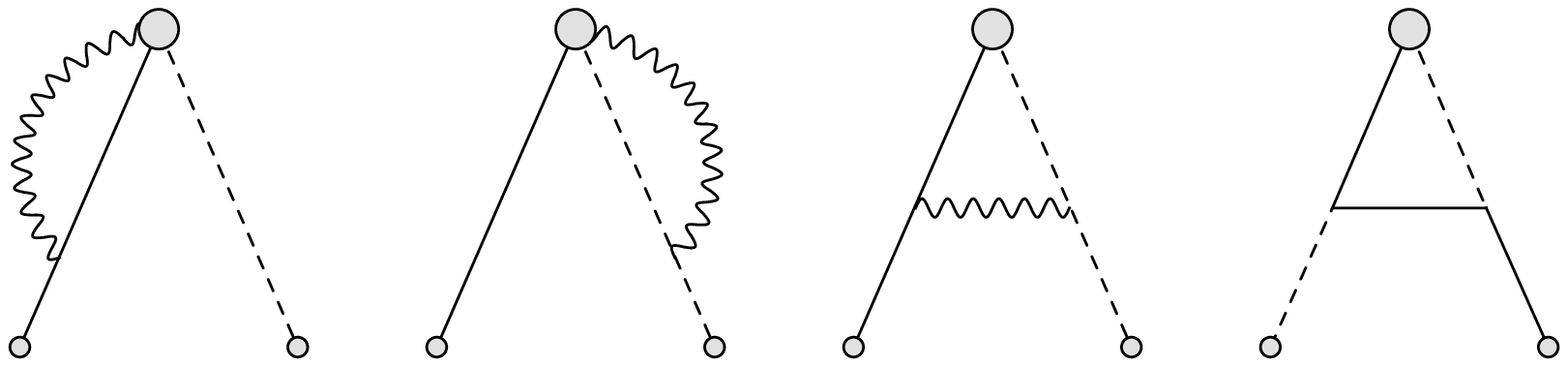}                     
    \caption{One-loop 1PI Feynman diagrams contributing to  $\l_k Q_{n-k} \lra Q_{k}\lh_{n-k} $ mixing.}
    \label{DiagrVH}
    \end{center}
    \end{figure}

The action of the Hamiltonian is
\bea \label{action1}
H'_{12} \l_k Q_{n-k} & = &2\sum^n_{k^{\pr}=0}a_{n,k,k'}\l_{k'} Q_{n-k'}
+2\sum^{n}_{k'=0}b_{n,k,k'} Q_{k'}\lh_{n-k'}\, , \\ \label{action2}
H'_{12} Q_{k}\lh_{n-k}& =& 2\sum^n_{k^{\pr}=0}\check{a}_{n,k,k'}Q_{k'}\lh_{n-k'}
+2\sum^{n}_{k'=0}\check{b}_{n,k,k'} \l_{k'} Q_{n-k'} \, ,
\eea
where
\bea 
\label{an}
a_{n,k,k'} & = & \delta_{k=k'}\left(h(k+1)-\frac{1}{2(n-k+1)}\right)
-\frac{\delta_{k\neq k^\pr}}{|k-k^\pr|}+\frac{\delta_{k<k'}}{k'+1}
\\ \nn
& & +\delta_{k=k'}\frac{1+\kappa^2}{4}\left(h(n-k)+h(n-k+1)\right)\, ,
\\
b_{n,k,k'} & = &-\kappa \frac{\delta_{k \geq k'}}{n-k'+1}\, , \\
 \label{can}
\check{a}_{n,k,k'} & = & \kappa^2 \delta_{k=k'}\left(h(n-k+1)-\frac{1}{2(k+1)}\right)
-\kappa^2 \frac{\delta_{k\neq k^\pr}}{|k-k^\pr|}+\kappa^2 \frac{\delta_{k>k'}}{n-k'+1}
\\\nn
& &+\delta_{k=k'}\frac{1+\kappa^2}{4}\left(h(k)+h(k+1)\right)\, ,
\\
\check{b}_{n,k,k'} & = &-\kappa\frac{\delta_{k \leq k'}}{k'+1} \,.
\eea
In this case, the Hamiltonian density $H'_{12}$ is {\it not} an $SU(1,1)$ invariant. However,
conformal symmetry only dictates that the total Hamiltonian $\sum_{\ell} H'_{\ell, \ell+1}$ acting on a {\it closed} spin chain must be
invariant. A redefinition of the two-site Hamiltonian of the form
\be \label{redefinition}
H'_{\ell, \ell + 1} \to H'_{\ell, \ell + 1} - K_\ell + K_{\ell +1}\, ,
\ee
where $K_\ell$ is a local operator at site $\ell$, leaves the total Hamiltonian invariant.  So what we must really
check is whether we can make the two-site Hamiltonian invariant by an appropriate choice of $K_\ell$. The choice of $K_\ell$
that makes $H'_{12}$ invariant  for the whole $SU(1,1) \times SU(1|1) \times SU(1|1) \times U(1)$ subsector is
\be \label{Kl}
K_\ell = \sum_{k=0}^\infty  \left( f(k)  P^\ell_{Q_k} -   f(k)  P^\ell_{\bar Q_k} \right) \, ,\quad f(k) = \frac{1-\kappa^2}{2}\left(h(k)+h(k+1)\right) \, ,
\ee
where $P^\ell_{Q_{k}}$ is the projector on the state $Q_k$ at site $\ell$, and similarly for $P^\ell_{\bar Q_{k}}$. We have verified this claim
for the restriction of $H'_{12}$ to each of the tensor products. For the tensor products $\Vm \times \Hm \leftrightarrow \Hm \times {\Vm}$, the transformation (\ref{redefinition}, \ref{Kl}) amounts
to redefining the coefficients (\ref{an}, \ref{can}) as
\be
a_{n, k, k'} \to  a_{n, k, k'} + \frac{1}{2}f(n-k)\, , \quad  \check a_{n, k, k'} \to \check a_{n, k, k'} -  \frac{1}{2}f(k)\,.
\ee
The new coefficients read
\bea 
a_{n,k,k'} & = & \delta_{k=k'}(h(k+1)+h(n-k))-\frac{\delta_{k\neq k'}}{|k-k'|}+\frac{\delta_{k<k'}}{k'+1}\, ,
\\
\check{a}_{n,k,k'} & = & \kappa^2 \left(\delta_{k=k'}(h(k)+h(n-k+1))-\frac{\delta_{k\neq k'}}{|k-k'|}+\frac{\delta_{k>k'}}{n-k'+1} \right)\, ,
\eea
and these combinations \textit{are} $SU(1,1)$ invariant as can be easily checked with the oscillator representation. (The coefficients $b_{n,k,k'}$ and $\check{b}_{n,k,k'}$ were never problematic).
Now we can write $H'_{12}$ in (\ref{action1}, \ref{action2})
as a sum of projectors
\be
H'_{12} = \sum_{j=0}^{\infty} 
\left(
\begin{array}{cc}
A_{11}(j)& A_{12}(j)\\
A_{21}(j) & A_{22}(j)\\
\end{array}\right)\Pm'_{-\frac{3}{2}-j} \,.
\ee
To obtain the undetermined coefficients we act on the $SU(1,1)$ highest weights (of spin $-\frac{3}{2}-j$),
\bea 
\Jm(j) & = & \sum_{k=0}^j(-1)^k\binom{j}{k}\binom{j+1}{k}\Dm^{j-k}\l_{2+} \Dm^k Q_2\, ,
\\
\Km(j) & = & \sum_{k=0}^j(-1)^k\binom{j}{k}\binom{j+1}{k+1}\Dm^{j-k}Q_2 \Dm^k \lh_{2+}\, .
\eea
As before, these are also superconformal descendants. They can be obtained by applying 
 $-\frac{1}{2}\Rm_{2}^{\ph{2}1}\Rm_{2}^{\ph{2}1}\Qm_{+}^{\ph{+}2}$ to (\ref{VHprimaryqm1}) and  (\ref{HVprimaryqm1})  for $j=0$ 
and $\Qm_{+}^{\ph{+}1}\tilde{\Qm}_{\dot{+}2}$ to (\ref{VHprimaryq}) and (\ref{HVprimaryq}) for $j>0$.
\bea
H'_{12}\Jm(j) & = & 2\left(h(j+1)+h(j)\right)\Jm(j)-\frac{2\kappa}{j+1}\Km(j)\, ,
\\
H'_{12}\Km(j) & = & 2\kappa^2 \left(h(j+1)+h(j)\right)\Km(j)-\frac{2\kappa}{j+1}\Jm(j)\, .
\eea
The lifting procedure works as before: there is a one-to-one relationship between $SU(1,1)$ modules and superconformal modules, 
now with $q+1=j$. The full one-loop result for  $\Vm \times \Hm \leftrightarrow \Hm \times {\Vm}$  is then
\be
\label{FullHonVxH}
H_{12} = 2\sum_{q=-1}^{\infty} 
\left(
\begin{array}{cc}
h(q+2)+h(q+1)& -\frac{\kappa}{q+2}\\
-\frac{\kappa}{q+2} & \kappa^2(h(q+2)+h(q+1))\\
\end{array}\right)\Pm_{(\frac{q+1}{2},\frac{q}{2})}\, .
\ee
\textit{A quick check:} Let's consider the action of the Hamiltonian on the two dimensional vector space formed by $\f Q$ and $Q \fh$. These are the superconformal primaries of the $q=-1$ modules. The mixing matrix is just (\ref{FullHonVxH}) evaluated at $q=-1$. The result is
\be 
H_{12}=\left(
\begin{array}{cc}
2  & -2\kappa\\
-2\kappa & 2\kappa^2  \\
\end{array}\right)\, ,
\ee
in perfect agreement with \cite{Gadde:2010zi}. This is a nice check because in the above calculation we never considered $\phi$ and 
$\fh$.

\subsection{$\Hm \times \Hm \leftrightarrow \Vm \times \bar{\Vm} \leftrightarrow \bar{\Vm} \times \Vm$}
\label{HxHFeynman}

The relevant states are $Q_k\bar{Q}_{n-k}$, $\l_{k }\bar{\l}_{n-k}$ and $\bar{\l}_{k} \l_{n-k}$. 

\begin{figure}[h]
    \begin{center}
    \includegraphics[scale=0.6]{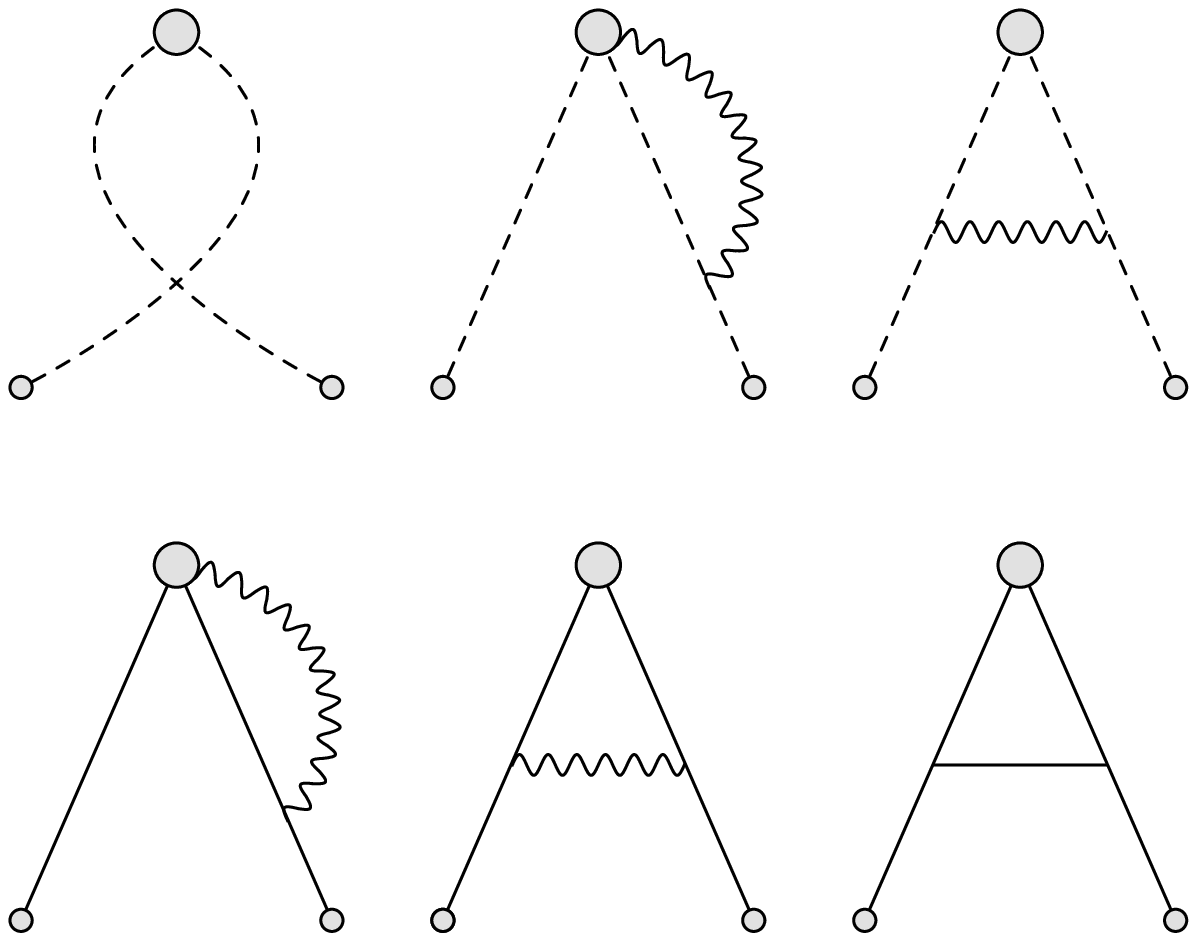}                     
    \caption{The diagrams in the first row contribute to $Q_k\bar{Q}_{n-k}$ mixing, the diagrams in the second row to $\l_k \bar{\l}_{n-k}$ and 
    $\bar{\l}_k \l_{n-k}$ mixing.}
    \label{DiagrHHVV1}
    \end{center}
    \end{figure}
      \begin{figure}[h]
    \begin{center}
    \includegraphics[scale=0.6]{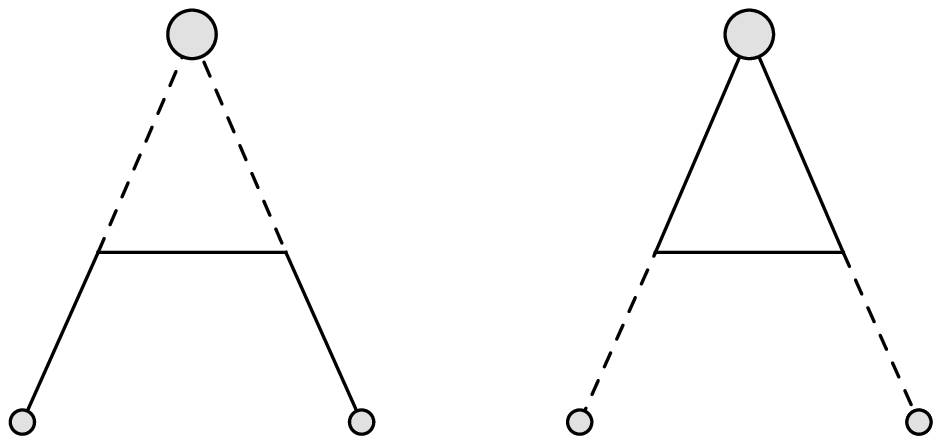}                     
    \caption{1PI diagrams contributing to the mixing $Q_k \bar{Q}_{n-k} \lra (\l_k \bar{\l}_{n-k},\bar{\l}_k\l_{n-k})$}
    \label{DiagrHHVV2}
    \end{center}
    \end{figure}
The action of the Hamiltonian on the squarks, {\it after} the redefinition  (\ref{redefinition}, \ref{Kl}) to make it $SU(1,1)$ invariant, is
\be 
	H'_{12} Q_{k\, \hat{\Im}} \bar{Q}_{n-k}^{\hat{\Jm}}=2\sum^n_{k'=0}\left( a_{n,k,k'}\right)_{\hat {\cal I} \hat {\cal L}}^{\hat {\cal J} \hat {\cal K}}\, Q_{k'\, \hat{\Km}}\bar{Q}_{n-k'}^{\hat{\Lm}}
	+2\delta_{\hat{\Im}}^{\hat{\Jm}}  \sum^{n-1}_{k'=0}\left(b_{n,k,k'}\l_{k'}\bar{\l}_{n-k'-1}
	+c_{n,k,k'}\bar{\l}_{k'}\l_{n-k'-1}
	\right) \, ,
	\ee
where (with $\mathbb{\hat{I}} \equiv \delta_{\hat{\Im}}^{\hat{\Km}} \delta^{\hat{\Jm}}_{\hat{\Lm}}$, $\mathbb{\hat{K}}\equiv \delta_{\hat{\Im}}^{\hat{\Jm}} \delta^{\hat{\Km}}_{\hat{\Lm}}$)
\bea
\label{ankQQllbFeyn}
a_{n,k,k^\pr} & = & \frac{\mathbb{\hat{K}}}{(n+1)}+ \kappa^2\mathbb{\hat{I}}\left(\delta_{k=k^\pr}(h(k)+h(n-k)) -\frac{\delta_{k\neq k^\pr}}{|k-k^\pr|} \right)\, ,
\\
\label{bnkQQllbFeyn}
b_{n,k,k^{\pr}} & = & \frac{1}{n+1}\left(-\frac{\delta_{k>k^{\pr}}}{n-k^\pr}+\frac{\delta_{k\leq k^{\pr}}}{k^\pr +1}\right)\, ,
\\
\label{cnkQQllbFeyn}
c_{n,k,k^{\pr}} & = &  -\frac{1}{n+1}\left(-\frac{\delta_{k>k^{\pr}}}{n-k^\pr}+\frac{\delta_{k\leq k^{\pr}}}{k^\pr +1}\right)\,.
\eea
For the action on the fermions, we get 
\bea
\nn
H_{12}\l_k \bar{\l}_{n-k} & = & 2\sum^n_{k'=0}\left(a_{n,k,k'}\l_{k^{\pr}}\bar{\l}_{n-k^{\pr}}
+b_{n,k,k^{\pr}}\bar{\l}_{k^{\pr}}\l_{n-k^{\pr}}\right)
+2\sum^{n+1}_{k^{\pr}=0}c_{n,k,k^{\pr}}Q_{k'\, \hat{\Im}}\bar{Q}_{n+1-k'}^{\hat{\Im}}\, ,\\
\\
\nn
H_{12}\bar \l_k {\l}_{n-k} & = & 2\sum^n_{k'=0}\left(a_{n,k,k'}\bar \l_{k^{\pr}} {\l}_{n-k^{\pr}}
+b_{n,k,k^{\pr}} {\l}_{k^{\pr}} \bar \l_{n-k^{\pr}}\right)
-2\sum^{n+1}_{k^{\pr}=0}c_{n,k,k^{\pr}}Q_{k'\, \hat{\Im}}\bar{Q}_{n+1-k'}^{\hat{\Im}}\, , 
\\
\eea
where
\bea
\nn
a_{n,k,k^\pr} & = & \delta_{k=k^\pr}\left(h(k+1)+h(n-k+1)-\frac{1}{n+2}\right)-\frac{\delta_{k\neq k^{\pr}}}{|k-k^{\pr}|}
\\
\label{ankllbQQFeyn}
& &+\delta_{k>k^{\pr}}\frac{k+1}{(n+2)(n-k^\pr +1)}+\delta_{k<k^{\pr}}\frac{n-k+1}{(n+2)(k^\pr +1)}\, ,
\\
\label{bnkllbQQFeyn}
b_{n,k,k^{\pr}} & = & \frac{1}{n+2}\left(\delta_{k=k^\pr}+\delta_{k>k^{\pr}}\frac{n-k +1}{n-k^\pr +1}
+\delta_{k<k^{\pr}}\frac{k+1}{k^\pr +1} \right)\, ,
\eea
\bea
\label{cnkllbQQFeyn}
c_{n,k,k^{\pr}} & = & -\left((n-k+1)\left((k+1)b_{n+1,k+1,k^{\pr}}-(k+2)b_{n+1,k,k^{\pr}}\right)+k^\pr b_{n,k,k^{\pr}-1}\right)\,.
\eea
Let us now distinguish  the two possible combinations of $SU(2)_L$ indices:

\subsubsection{$SU(2)_L$ singlet}

The Hamiltonian density  can be written as
\be 
H'_{12} = A_{11}(0)\Pm'_{-1}+\sum_{j=1}^{\infty} 
\left(
\begin{array}{ccc}
A_{11}(j) & A_{12}(j) & A_{13}(j)\\
A_{21}(j) & A_{22}(j) & A_{23}(j) \\
A_{31}(j) & A_{32}(j) & A_{33}(j) \\
\end{array}\right)\Pm'_{-1-j}\, ,
\ee
To fix the undetermined constants we consider  the $SU(1,1)$ highest weights (of spin $-1-j$),
\bea
\Jm(j) & = & -\sum_{k=0}^{j}(-1)^k\binom{j}{k}\binom{j}{k} \ \Dm^{j-k}Q_2 \Dm^{k}\bar{Q}_2\, ,
\\
\Km(j)& = & \sqrt{2j(j+1)}\sum_{k=0}^{j-1}(-1)^k\frac{1}{k+1}\binom{j}{k}\binom{j-1}{i}\Dm^{j-k-1}\l_{2+} \Dm^{k}\bar{\l}_{2\dot{+}}\, ,
\\
\bar{\Km}(j) & = & -\sqrt{2j(j+1)}\sum_{k=0}^{j-1}(-1)^k\frac{1}{k+1}\binom{j}{k}\binom{j-1}{k}\Dm^{j-k-1}\bar{\l}_{2\dot{+}} \Dm^{k}\l_{2+} \,.
\eea
These states are superconformal descendants obtained by acting with
 $-\frac{1}{2}\Rm_{2}^{\ph{2}1}\Rm_{2}^{\ph{2}1}$ on (\ref{HHprimaryqm1}) for $j=0$,
and with $\Qm_{+}^{\ph{+}1}\tilde{\Qm}_{\dot{+}2}$ on (\ref{HHprimaryq}) and (\ref{VVbprimaryq}) 
for $j>0$.
The action of the Hamiltonian is, for $j >0$,
\bea 
\nn
H'_{12}\Jm(j) & = & 4\kappa^2 h(j)\Jm(j) + \frac{2\sqrt{2}}{ \sqrt{j(j+1)}} \Km(j) + \frac{2\sqrt{2}}{ \sqrt{j(j+1)}}\bar{\Km}(j)\, ,
\\
\nn
H'_{12}\Km(j) & = & 2(h(j+1)+h(j-1))\Km(j) - \frac{2}{j(j+1)}\bar{\Km}(j)+\frac{2\sqrt{2}}{ \sqrt{j(j+1)}}\Jm(j) \, ,
\\
\nn
H'_{12}\bar{\Km}(j) & = & 2(h(j+1)+ h(j-1))\bar{\Km}(j) - \frac{2}{j(j+1)}\Km(j)+\frac{2\sqrt{2}}{ \sqrt{j(j+1)}}\Jm(j)\, ,
\eea
and for $j=0$,
\be 
H'_{12}\Jm(0)  =  4\Jm(0)\, .
\ee
We can immediately read off  the full one-loop Hamiltonian density in the  $\Hm \times \Hm \leftrightarrow \Vm \times \bar{\Vm} \leftrightarrow \bar{\Vm} \times \Vm$ subspace,
\be 
\label{FullHonHxHs}
H_{12} = 4\Pm_{(-\frac{1}{2},-\frac{1}{2})}+2\sum_{q=0}^{\infty} 
\small{ \left(
\begin{array}{ccc}
2\kappa^2 h(q+1)&\frac{\sqrt{2}}{ \sqrt{(q+1)(q+2)}}  & \frac{\sqrt{2}}{ \sqrt{(q+1)(q+2)}} \\
\frac{\sqrt{2}}{ \sqrt{(q+1)(q+2)}}  & h(q+2)+ h(q) & - \frac{1}{(q+1)(q+2)} \\
\frac{\sqrt{2}}{ \sqrt{(q+1)(q+2)}}  & - \frac{1}{(q+1)(q+2)} & h(q+2)+ h(q) \\
\end{array}\right)\Pm_{(\frac{q}{2},\frac{q}{2})}} \,.
\ee
\textit{A quick check:} Let's consider the action of the Hamiltonian on the three-dimensional vector space spanned by $2\f \bar{\f}$, $2\bar{\f} \f$ and $Q_{\Im\,\hat{\Im}}\bar{Q}^{\hat{\Im}\,\Im}$. These are the superconformal primaries of the $q=0$ modules. The mixing matrix is the one given in (\ref{FullHonHxHs}) evaluated at $q=0$,
\be 
H_{12}=
\left(
\begin{array}{ccc}
4\kappa^2 & 2 & 2 \\
2 & 3  & -1\\
2 & -1 & 3  \\
\end{array}\right) \, ,
\ee
again in  agreement with \cite{Gadde:2010zi}. 

\subsubsection{$SU(2)_L$ triplet}

In this case  $\Hm \times \Hm$ does not mix with $\Vm \times \bar{\Vm}$ and $\bar{\Vm} \times \Vm$, and  the Hamiltonian on $\Hm \times \Hm$
is simply
\be 
\label{FullHonHxHt}
H_{12} = \sum_{q=0}^{\infty}4\kappa^2h(q+1)\Pm_{(\frac{q}{2},\frac{q}{2})}\,.
\ee

\section{Algebraic Evaluation of the Hamiltonian}
\label{algebraicS}

	In addition to the $SU(1,1)$ symmetry already exploited in the previous section, our closed subsector has an extra $SU(1|1) 
	\times SU(1|1)\times U(1)$ symmetry.
	The generators of the two $SU(1|1)$s are
	\be
	B=\frac{1}{2}\Lm_{-}^{\ph{-}-}+\frac{1}{2}\dot{\Lm}_{\dot{-}}^{\ph{\dot{-}}\dot{-}}+\frac{1}{2}D_0+r\, , \quad  \Sm(g) = 
	S_{1}^{\ph{1}-}(g)\, , \quad 
	\Qm(g) = Q_{-}^{\ph{-}1}(g)\, ,
	\ee
	\be
	\tilde{B}=\frac{1}{2}\Lm_{-}^{\ph{-}-}+\frac{1}{2}\dot{\Lm}_{\dot{-}}^{\ph{\dot{-}}\dot{-}}+\frac{1}{2}D_0-r\, , \quad 
	\tilde{\Sm}(g) 
	= 
	\tilde{S}^{\dot{-}2}(g)\, , \quad \tilde{\Qm}(g) = \tilde{Q}_{\dot{-}2}(g)\, ,
	\ee
	and can be checked to commute with the $SU(1,1)$ generators (\ref{SU(1,1)}). The $U(1)$ is a central element corresponding to the 
	quantum part 
	of the dilatation operator, $\delta D(g)$.
	
	The (anti)commutators are
	\begin{align}
	 [B,\Qm(g) ] & = \Qm(g)\, , & 
	 [\tilde{B},\tilde{\Qm}(g) ] & = \tilde{\Qm}(g)\, ,
	 \\
	 \lbrack B,\Sm(g) ] & =  -\Sm(g)\, ,
	 &\lbrack \tilde{B},\tilde{\Sm}(g) ] & =  -\tilde{\Sm}(g)\, ,
	 \\ \label{QS}
	 \{\Sm(g),\Qm(g)\} & =  \frac{1}{2}\delta D(g)\, , & 
	  \{\tilde{\Sm}(g),\tilde{\Qm}(g)\} & =  \frac{1}{2}\delta D(g) \,.
	\end{align}
The operator $L=B+\tilde{B}$ evaluates to 1 on each of the elementary letters of the subsector, and thus measures
the ``length'' of a state. Since 
	\begin{align}
	 [L,\Qm(g) ] & =  \Qm(g)\, , & 
	  [L,\tilde{\Qm}(g) ] & = \tilde{\Qm}(g) \, , & 
	 \\
	 [L,\Sm(g) ] & =  -\Sm(g)\, , & 
	  [L,\tilde{\Sm}(g) ] & = -\tilde{\Sm}(g) \, . & 
	\end{align}
we learn that $\Qm(g)$ and $\tilde{\Qm}(g)$ increase the length of a state by one unit while $\Sm(g)$ and $\tilde{\Sm}(g)$ decrease it.
	
	\subsection{First order expressions for $\Qm(g)$ and $\Sm(g)$}
	
In the classical limit $g \to 0$ one easily checks that the $SU(1|1)$ generators annihilate all the states of the subsector, consistent
with the fact that they must	 change the length of a state. As in \cite{Beisert:2004ry}, we know that there must be quantum corrections
to $\Qm(g)$ and $\Sm(g)$, because their anticommutator must yield a non-vanishing quantum dilation operator.
Writing $\Qm(g) = g \Qm + O(g^2)$, the most general ansatz for the action of $\Qm$ on $\lambda$ compatible
with Lorentz and R-charge conservation is
	\bea 
	\nn
	\Qm \l_n & = & \sum_{k'=0}^n a_{n,k'} Q_{k'}\bar{Q}_{n-k'} 
	\\
	& & + \sum_{k'=0}^{n-1} b_{n,k'} \l_{k'}\bar{\l}_{n-k'-1} + \sum_{k'=0}^{n-1} c_{n,k'} \bar{\l}_{k'}\l_{n-k'-1}
	\eea
	for arbitrary coefficients $a_{n,k'}$, $b_{n,k'}$ and $c_{n,k'}$. The coefficients can be constrained by requiring
	that $\Qm$ commutes with the $SU(1,1)$ 
	algebra. Requiring $[\Jm', \Qm] \l_n=0$
	 fixes $a_{n,k'}$ to be a constant $a_{n,k'}=\a'$, and  $b_{n,k'} = c_{n,k'} = 0$. This is however too restrictive,
	and as in  $\Nm=4$ SYM \cite{Beisert:2004ry}, one should only require that $[\Jm', \Qm]$ annihilates all gauge invariant
	states (closed spin chains). We should demand
	$[\Jm', \Qm] \l_n \sim 0$, where $\sim$ stands for equivalence up to a gauge transformation.
	 There are two independent gauge transformations, corresponding to adding an extra  $\bar{\l}$ or $\bar{\lh}$ to the chain, so we impose
	\bea 
	[\Qm, \Jm'_+] \l_n & = & \a \left( \l_n \bar{\l} + \bar{\l}\l_n \right)\, ,
	\\
	\lbrack \Qm, \Jm'_+] \bar{\l}_n & = &\a \left( \bar{\l}_n \bar{\l} + \bar{\l}\bar{\l}_n \right)\, ,
	\\
	\lbrack \Qm, \Jm'_+] Q_n & = & \a \left( \bar{\l} Q_n - \g Q_n \bar{\lh} \right)\, ,
	\\
	\lbrack \Qm, \Jm'_+] \bar{Q}_n & = & \a \left( \g \bar{\lh} \bar{Q}_n - \bar{Q}_n \bar{\l} \right)\, ,
	\\
	\lbrack \Qm, \Jm'_+] \lh_n & = & \a \g \left( \lh_n \bar{\lh} +  \bar{\lh}\lh_n \right)\, ,
	\\
	\lbrack \Qm, \Jm'_+] \bar{\lh}_n & = & \a \g \left( \bar{\lh}_n \bar{\lh} + \bar{\lh}\bar{\lh}_n \right)\,,
	\eea
	where we have labelled by $\alpha$ and $\alpha \gamma$ the two independent gauge parameters.
	We now find 
	\bea 
	a_{n, k'} & = & \alpha' \, ,
	\\
	b_{n,k'} & = & \frac{\a}{n-k'}\, ,
	\\
	c_{n,k'} & = & \frac{\a}{k'+1}\, ,
	\eea
	where at this stage $\a$  and $\alpha'$ are arbitrary constants.
    Similarly, for the action on the other states of the sector,
	\bea 
	\nn
	\Qm \lh_n & = & \sum_{k'=0}^n \a'' \bar{Q}_{k'} Q_{n-k'}
	\\
	& & + \a \g\left(\sum_{k'=0}^{n-1} \frac{1}{n-k'} \lh_{k'}\bar{\lh}_{n-k'-1} + \sum_{k'=0}^{n-1} \frac{1}{k'+1}
	\bar{\lh}_{k'}\lh_{n-k'-1} \right)\, ,
	\\
	\Qm \bar{\l}_n & = & \a \sum_{k'=0}^{n-1}\frac{n+1}{(k'+1)(n-k')}\bar{\l}_{k'} \bar{\l}_{n-k'-1}\, ,	
	\\
	\Qm \bar{\lh}_n & = & \a \g \sum_{k'=0}^{n-1}\frac{n+1}{(k'+1)(n-k')}\bar{\lh}_{k'} \bar{\lh}_{n-k'-1}\, ,
	\\
	\Qm Q_n & = & \a \sum_{k'=0}^{n-1}\left( \frac{1}{k'+1}\bar{\l}_{k'} Q_{n-k'-1}
	-\frac{\g}{n-k'}Q_{k'} \bar{\lh}_{n-k'-1}\right)\, ,
	\\
	\Qm \bar{Q}_n & = & \a \sum_{k'=0}^{n-1}\left( \frac{\g}{k'+1}\bar{\lh}_{k'} \bar{Q}_{n-k'-1}
	-\frac{1}{n-k'}\bar{Q}_{k'} \bar{\l}_{n-k'-1}\right)\, .
	\eea
One can check that the  commutators $[\Jm'_-,\Qm] = 0$ and  $[\Jm'_3,\Qm]  = 0$ are then identically satisfied with the action of $\Qm$ given by the above expressions.
	An analogous analysis can be performed for $\Sm$. Now
	the relevant gauge transformations are
	\begin{align}
	 \lbrack \Sm, \Jm'_-]\bar{\l}_{k} \bar{\l}_{n-k} & =  \b \left( \delta_{k=0} + \delta_{n=k} \right) \bar{\l}_{n}\, , & 
	  \lbrack \Sm, \Jm'_-]\bar{\lh}_{k} \bar{\lh}_{n-k} & =  \b\g'\left( \delta_{k=0} + \delta_{n=k} \right) \bar{\lh}_{n}\, ,
	 \\
	 [\Sm, \Jm'_-]\l_k \bar{\l}_{n-k} & = \b \delta_{n=k} \l_n\, ,  & [\Sm, \Jm'_-]\lh_k \bar{\lh}_{n-k} & =  \b \g' \delta_{n=k} \lh_n\, , 
	\\
	 \lbrack \Sm, \Jm'_-]\bar{\l}_{k} \l_{n-k} & = \b \delta_{k=0} \l_{n}\, , & 
	 \lbrack \Sm, \Jm'_-]\bar{\lh}_{k} \lh_{n-k} & = \b \g' \delta_{k=0} \lh_{n}\, ,
	\\
	\lbrack \Sm, \Jm'_-]\bar{\l}_{k} Q_{n-k} & = \b \delta_{k=0} Q_{n}\, , & 
	\lbrack \Sm, \Jm'_-]\bar{\lh}_{k} \bar{Q}_{n-k} & = \b \g'\delta_{k=0} \bar{Q}_{n}\, ,
	\\
	 \lbrack \Sm, \Jm'_-]\bar{Q}_{k} \bar{\l}_{n-k} & =  - \b \delta_{n=k} \bar{Q}_{n}\, , &  
	 \lbrack \Sm, \Jm'_-] Q_{k} \bar{\lh}_{n-k} & =  - \b \g' \delta_{n=k} Q_{n}\, ,
	\end{align}
	and the action of $\Sm$ consistent with them is
	\begin{align}
	 \Sm Q_{k\hat{\Im}} \bar{Q}_{n-k}^{\hat{\Jm}} & =  \frac{\b'}{n+1}\l_n \delta_{\hat{\Im}}^{\hat{\Jm}}\, , & 
	 \Sm \bar{Q}_{k}^{\hat{\Jm}} Q_{n-k\hat{\Im}}  & = \frac{\b''}{n+1}\lh_n \delta_{\hat{\Im}}^{\hat{\Jm}}\, ,
	 \\
	 \Sm \bar{\l}_k \bar{\l}_{n-k} & =  \b \bar{\l}_{n+1}\, , & \Sm \bar{\lh}_k \bar{\lh}_{n-k} & = \g' \b \bar{\lh}_{n+1} \, ,
	 \\
	 \Sm \l_k \bar{\l}_{n-k} & =  \b \frac{k+1}{n+2}\l_{n+1}\, , & 
	 \Sm \lh_k \bar{\lh}_{n-k} & =  \g' \b \frac{k+1}{n+2}\lh_{n+1}\, ,
	 \\
	 \Sm \bar{\l}_k \l_{n-k} & =  \b \frac{n-k+1}{n+2}\l_{n+1}\, , & 
	 \Sm \bar{\lh}_k \lh_{n-k} & = \g' \b \frac{n-k+1}{n+2}\lh_{n+1}\, ,
	 \\
	 \Sm \bar{\l}_k Q_{n-k} & =  \b Q_{n+1}\, , & \Sm Q_{k} \bar{\lh}_{n-k} & = -\g' \b Q_{n+1}\, , 
	 \\
	 \Sm \bar{Q}_k \bar{\l}_{n-k} & =  -\b \bar{Q}_{n+1}\, , & \Sm Q_{k} \bar{\lh}_{n-k} & = -\g' \b Q_{n+1}\, .
	 	\end{align}
	With these expressions, the remaining commutators $[\Jm'_+, \Sm] = 0$ and  $[\Jm'_3,\Sm]  = 0$ are automatically satisfied.

	As we are interested in unitary representations of the superconformal algebra,  we  impose the hermiticity condition\footnote{To exhibit hermiticity explicitly
	one needs to  rescale the fermion letters as $\chi_n \to\frac{\chi_n}{\sqrt{n+1}}$, where $\chi_n$ stands for $\lambda_n\, ,\check \lambda_n\,, \bar {\lambda}_n$ or $\bar {\check \lambda}_n$.}
	\be 
	\Qm^{\dag}  =  \Sm \, ,
	\ee
	which implies  the following reality constraints for the undetermined coefficients:
	\begin{align}
	\a & = \b^*\, ,
	\\
	\a' & =	\b'^*\, ,
	\\
	\a'' & = \b''^*\, ,
	\\
	\g & = \g'^* \, .
	\end{align}	
	Having determined the $O(g)$ action of ${\cal Q}(g)$ and ${\cal S}(g)$, we
	 are now in the position to evaluate the one-loop Hamiltonian,  since  the algebra (\ref{QS}) implies
	\be 
	H' = 2 \{\Sm,\Qm\} \, .
	\ee
	Let us proceed to find $H'$ in the different subspaces:

 \subsection{ $\Vm \times \Vm$ and $\bar{\Vm} \times \bar{\Vm}$ }
 
 	The $\bar{\Vm} \times \bar{\Vm}$ case is identical with $\Nm=4$, we refer the interested reader to \cite{Beisert:2004ry} 
	for details of the calculation. The result is
	\be
	H'_{12}\bar{\l}_k \bar{\l}_{n-k} = 2|\a|^2 \sum_{k'=0}^{n} c_{n,k,k'} \bar{\l}_{k'} \bar{\l}_{n-k'}\, ,
	\ee
	with
	\be 
	c_{n,k,k'}=\delta_{k=k'}\left(h(k+1)+h(n-k+1) \right) -\frac{\delta_{k\neq k'}}{|k-
	k'|}+\frac{\delta_{k>k'}}{n-
	k'+1}+\frac{\delta_{k< k'}}{k'+1}\, .
	\ee
	For $\Vm \times \Vm$ the calculation is very similar, and the result is
	\be
	H'_{12}\l_k \l_{n-k} = 2|\a|^2 \sum_{k'=0}^{n} c_{n,k,k'} \l_{k'} \l_{n-k'}\, ,
	\ee
	with
	\be 
	c_{n,k,k'}=\delta_{k=k'}\left(h(k+1)+h(n-k+1) +\frac{|\a'|^2}{|\a|^2} - 1\right) -\frac{\delta_{k\neq k'}}{|k-
	k'|}+\frac{\delta_{k>k'}}{n-
	k'+1}+\frac{\delta_{k< k'}}{k'+1} \,.
	\ee
	We now impose the physical requirement that the action of the Hamiltonian on $\lambda \lambda$ is identical to the action on $\bar \lambda \bar \lambda$
	(this is CPT invariance in the field theory). 
	 This fixes $|\a'|^2 = |\a|^2$, which implies $\a'=e^{i\theta_1}\a$, where $\theta_1$ is an arbitrary phase. Proceeding just as
	 in Section \ref{VxVFeynman} we can uplift the result to the full theory,
	\be 
	H_{12}= 0 \times \Pm_{\bar{\Em}}+|\a|^2\sum_{q=0}^{\infty}4h(q+1)\Pm_{(\frac{q+1}{2},\frac{q-1}{2})}\, .
	\ee
	The overall constant $|\a|^2$ cannot be fixed algebraically and is related to a rescaling of the coupling. To match with the
	field theory result (\ref{FullHonVxV}) we need to set $|\a|^2 = 1$.
	
	\subsection{$\bar{\Vm} \times \Hm \lra \Hm \times \bar{\Vm}$}
	
	Since this case is somewhat different from  $\Nm=4$ SYM because of  multiplet mixing, let us give a few more details of the calculation.
We need to evaluate
	\be 
	H'_{12} \l_k Q_{n-k}= 2 (\Sm\Qm + \Qm\Sm) \l_k Q_{n-k}\, .
	\ee
	In the first term inside the parenthesis we can act with $\Qm$ in either the first or the second site, we will denote this 
	contributions by $\Qm_1$ and $\Qm_2$. Both choices will increase the length of the chain by one, which implies that $\Sm$ can 
	act in either sites 1-2 or  
	2-3, we will denote this by $\Sm_{12}$ and $S_{23}$. Taking into account all possible combinations the action of the 
	Hamiltonian is 
	\be 
	H'_{12} = 2\left(\Sm_{12}\Qm_1	+ \Sm_{23}\Qm_1	+ \Sm_{12}\Qm_2	+ \Sm_{23}\Qm_2	+ \Qm_1	\Sm_{12}\right).
	\ee
	Each individual contribution can be calculated by straightforward application of the action of the supercharges given in the 
	previous section,
	\bea 
	\Sm_{12}\Qm_1 \bar{\l}_k Q_{n-k} & = & 4 h(k) \bar{\l}_k Q_{n-k}\, ,
	\\
	\Sm_{23}\Qm_1 \bar{\l}_k Q_{n-k} & = & - 2\sum_{k'=0}^{k-1}\left( \frac{1}{k'+1} + \frac{1}{k-k'}\right)
	\bar{\l}_{k'} Q_{n-k'}\, ,
	\\
	\Sm_{12}\Qm_2 \bar{\l}_k Q_{n-k} & = & - 2\sum_{k'=k+1}^n \left( \frac{1}{k'-k}\bar{\l}_{k'} Q_{n-k'} 
	 + \frac{\g }{n-k'+1} Q_{k'} \bar{\lh}_{n-k'} \right)\, ,
	\\
	\Sm_{23}\Qm_2 \bar{\l}_k Q_{n-k} & = & 2(1+|\g|^2)h(n-k)\bar{\l}_k Q_{n-k}\, , 
	\\
	\Qm_1\Sm_{12} \bar{\l}_k Q_{n-k} & = & 2\sum_{k'=0}^n \left( \frac{1}{k'+1}\bar{\l}_{k'} Q_{n-k'} 
	 - \frac{\g}{n-k'+1} Q_{k'} \bar{\lh}_{n-k'} \right) \, .
	\eea
	Now, since $\Sm_{12}\Qm_1$ and $\Sm_{23}\Qm_2$ act at the single site level, they are analogous to the self-energy contributions in 
	the field theory calculation. As usual for spin chains, we distribute them evenly in two adjacent sites by adding 
	an extra factor of one half. An analogous calculation can be done for $2\{\Sm,\Qm\} Q_{k}\lh_{n-k}$, the action of the Hamiltonian in this subspace is
\bea 
H'_{12} \l_k Q_{n-k} & = &2\sum^n_{k^{\pr}=0}a_{n,k,k'}\l_{k'} Q_{n-k'}
+2\sum^{n}_{k'=0}b_{n,k,k'} Q_{k'}\lh_{n-k'}\, , \\ 
H'_{12} Q_{k}\lh_{n-k}& =& 2\sum^n_{k^{\pr}=0}\check{a}_{n,k,k'}Q_{k'}\lh_{n-k'}
+2\sum^{n}_{k'=0}\check{b}_{n,k,k'} \l_{k'} Q_{n-k'} \, ,
\eea
where
\bea 
a_{n,k,k'} & = & \delta_{k=k'}\left(h(k+1)+\frac{1+|\g|^2}{2}h(n-k)\right)
	-\frac{\delta_{k\neq k'}}{|k-k'|}+\frac{\delta_{k<k'}}{k'+1}\, ,
\\
b_{n,k,k'} & = &-\g \frac{\delta_{k \geq k'}}{n-k'+1} \, ,
\\
\nn
\check{a}_{n,k,k'} & = & \frac{1+|\g|^2}{2} h(k)\delta_{k=k'}
+|\g|^2\left(h(n-k+1)\delta_{k=k'}- \frac{\delta_{k\neq k^\pr}}{|k-k^\pr|}+\frac{\delta_{k>k'}}{n-k'+1}\right)\, ,
\\
\\
\check{b}_{n,k,k'} & = &-\g^*\frac{\delta_{k \leq k'}}{k'+1} \,.
\eea
	This expression for the two-site Hamiltonian has the same problem we encountered in the Feynman diagram calculation: it is not $SU(1,1)$ invariant. But
	it {\it can} be made invariant by performing the gauge transformation (\ref{Kl}), now with
	 	\be 
	f(k)=(1-|\g|^2)h(k) \,.
	\ee 
	The uplifting to the full theory works exactly as in Section \ref{VxHFeynman}.
	Defining $\g \equiv \eta e^{i \theta_2}$, where $\eta$ and $\theta_2$ are real parameters, we find
	\be
	H_{12} = 2\sum_{q=-1}^{\infty} 
	\left(
	\begin{array}{cc}
	h(q+2)+h(q+1)& -\frac{\eta}{q+2}e^{i\theta_2}\\
	-\frac{\eta}{q+2}e^{-i\theta_2} & \eta^2(h(q+2)+h(q+1))\\
	\end{array}\right)\Pm_{(\frac{q+1}{2},\frac{q}{2})} \,.
	\ee
	The phase $\theta_2$ does not enter in any physical anomalous dimension, and can in fact be set to zero by a similarity transformation.
	Comparison with (\ref{FullHonVxH}) shows then perfect agreement with the field theory calculation, if we identify $\eta \equiv \kappa$.

	\subsection{$\Hm \times \Hm \leftrightarrow  \Vm  \times \bar{\Vm}\leftrightarrow \bar{\Vm} \times \Vm$}

	Following similar steps as in the previous subsection,
	we obtain for  this subspace
\be 
	H'_{12} Q_{k\, \hat{\Im}} \bar{Q}_{n-k}^{\hat{\Jm}}=2\sum^n_{k'=0}\left( a_{n,k,k'}\right)_{\hat {\cal I} \hat {\cal L}}^{\hat {\cal J} \hat {\cal K}}\, Q_{k'\, \hat{\Km}}\bar{Q}_{n-k'}^{\hat{\Lm}}
	+\delta_{\hat{\Im}}^{\hat{\Jm}}  2\sum^{n-1}_{k'=0}\left(b_{n,k,k'}\l_{k'}\bar{\l}_{n-k'-1}
	+c_{n,k,k'}\bar{\l}_{k'}\l_{n-k'-1}
	\right) \, ,
	\ee
	with
\bea
a_{n,k,k^\pr} & = & \frac{\mathbb{\hat{K}}}{(n+1)}+ \kappa^2\mathbb{\hat{I}}\left(\delta_{k=k^\pr}(h(k)+h(n-k)) -\frac{\delta_{k\neq k^\pr}}{|k-k^\pr|} \right)\, ,
\\
b_{n,k,k^{\pr}} & = & \frac{ e^{-i\theta_1}}{n+1}\left(-\frac{\delta_{k>k^{\pr}}}{n-k^\pr}+\frac{\delta_{k\leq k^{\pr}}}{k^\pr +1}\right)\, ,
\\
c_{n,k,k^{\pr}} & = & - \frac{ e^{-i\theta_1}}{(n+1)}\left(-\frac{\delta_{k>k'}}{n-k'}+\frac{\delta_{k\leq k'}}{k' 
	+1}\right)\, .
\eea
This expressions precisely coincide with our previous Feynman diagram results (\ref{ankQQllbFeyn})-(\ref{cnkQQllbFeyn}) apart from the extra $e^{-i\theta_1} $ phases in the cross terms.
For the action on the fermions, we get 
\bea
\nn
H_{12}\l_k \bar{\l}_{n-k} & = & 2\sum^n_{k'=0}\left(a_{n,k,k'}\l_{k^{\pr}}\bar{\l}_{n-k^{\pr}}
+b_{n,k,k^{\pr}}\bar{\l}_{k^{\pr}}\l_{n-k^{\pr}}\right)
+2\sum^{n+1}_{k^{\pr}=0}c_{n,k,k^{\pr}}Q_{k'\, \hat{\Im}}\bar{Q}_{n+1-k'}^{\hat{\Im}}\, ,
\\
\\
\nn
H_{12}\bar \l_k {\l}_{n-k} & = & 2\sum^n_{k'=0}\left(a_{n,k,k'}\bar \l_{k^{\pr}} {\l}_{n-k^{\pr}}
+b_{n,k,k^{\pr}} {\l}_{k^{\pr}} \bar \l_{n-k^{\pr}}\right)
-2\sum^{n+1}_{k^{\pr}=0}c_{n,k,k^{\pr}}Q_{k'\, \hat{\Im}}\bar{Q}_{n+1-k'}^{\hat{\Im}}\, ,
\\
\eea
	where
	\bea
	\nn
	a_{n,k,k'} & = & \delta_{k=k'}\left(h(k+1)+h(n-k+1)-\frac{1}{n+2}\right)-\frac{\delta_{k\neq k'}}{|k-k'|}
	\\
	& &+\delta_{k>k'}\frac{k+1}{(n+2)(n-k'+1)}+\delta_{k<k'}\frac{n-k+1}{(n+2)(k'+1)}\, ,
	\\
	b_{n,k,k'} & = & \frac{ e^{i\theta_1}} {n+2}\left(\delta_{k=k'}+\delta_{k>k'}\frac{n-k +1}{n-k'+1}
	+\delta_{k<k'}\frac{k+1}{k'+1} \right)\, ,
	\\
	\label{cnkllbQQAlg}
	c_{n,k,k'} & = & - e^{i\theta_1} \left(-\delta_{k \ge k'} + \frac{k+1}{n+2}\right) \, ,
	\eea
	again in agreement with the Feynman diagram result (\ref{ankllbQQFeyn})-(\ref{cnkllbQQFeyn}) up to the extra $ e^{i\theta_1} $ 
	factors.
	(Note that $c_{n,k,k'}$ coefficients in (\ref{cnkllbQQFeyn}) and in (\ref{cnkllbQQAlg}) are equal, thanks to a non-trivial identity.)
		Uplifting to the full theory gives
	\be 
	\label{full}
	H_{12} = 4\Pm_{(-\frac{1}{2},-\frac{1}{2})}+2\sum_{q=0}^{\infty} 
	\left(
	\begin{array}{ccc}
	2\kappa^2 h(q+1)& \frac{\sqrt{2}}{ \sqrt{(q+1)(q+2)}} e^{-i\theta_1}  & \frac{\sqrt{2}}{ \sqrt{(q+1)(q+2)}} e^{-i\theta_1} \\
	\frac{\sqrt{2}}{ \sqrt{(q+1)(q+2)}}  e^{i\theta_1}  & h(q+2)+ h(q) & - \frac{1}{(q+1)(q+2)} \\
	\frac{\sqrt{2}}{ \sqrt{(q+1)(q+2)}}  e^{i\theta_1}  & - \frac{1}{(q+1)(q+2)} & h(q+2)+ h(q) \\
	\end{array}\right)\Pm_{(\frac{q}{2},\frac{q}{2} )} \, .
	\ee
	The phase $\theta_1$ can be set to zero by a similarity transformation, and we find perfect agreement with the field theory
	answer (\ref{FullHonHxHs}).

\section{The Harmonic Action}
\label{harmonicS}

While we have obtained an explicit expression for the full one-loop Hamiltonian in terms of superconformal projectors,
evaluating this expression on concrete states is still a  rather cumbersome procedure.
For $\Nm=4$ SYM Beisert \cite{Beisert:2003jj}  was able to find a very explicit and elegant formula
for the  the action of the Hamiltonian on any state, using the oscillator representation, which he called the ``harmonic action''.
  Beisert's approach easily generalizes to our case and allows to write a harmonic action for the interpolating SCFT.

\subsection{\textbf{$\Vm \times \Vm$}}

For a state in $\Vm \times \Vm$ we found  that the action of the Hamiltonian is identical with that of $\Nm=4$ SYM. Let's review then how the harmonic action works in this case. As pointed out in \cite{Beisert:2003jj} a general state in $\Vm \times \Vm$ can be written as 
\be 
|s_1,...,s_n;A\rangle_{\text{\tiny{$\Vm\times\Vm$}}} = A^\dag_{s_1 A_1}... A^\dag_{s_n A_n}|0\rangle \otimes |0\rangle\, ,
\ee
where $A^\dag_A=(\ab^\dag_{\a},\bb^\dag_{\ad},\cb^\dag_{\Im})$ and $s_i=1,2$ indicates in which site the oscillator sits. The action of the Hamiltonian on this state does not change the number of oscillators but merely shifts them from site 1 to site 2 (or vice versa) in all possible combinations. This can be written as
\be 
H_{12}|s_1,...,s_n;A\rangle_{\text{\tiny{$\Vm\times\Vm$}}} = \sum_{s'_1,...,s'_n}c_{n,n_{12},n_{21}}\delta_{C_1,0} \delta_{C_2,0}|s'_1,...,s'_n;A\rangle_{\text{\tiny{$\Vm\times\Vm$}}} \, ,
\ee
where the delta functions project onto states with zero central charge and $n_{ij}$ counts the number of oscillators moving from site $i$ to site $j$. The explicit formula for the function $c_{n,n_{12},n_{21}}$ is
\be 
c_{n,n_{12},n_{21}}=(-1)^{1+n_{12}n_{21}}\frac{\Gamma(\frac{1}{2}(n_{12}+n_{21}))\Gamma(1+\frac{1}{2}(n-n_{12}-n_{21}))}{\Gamma(1+\frac{1}{2}n)}\, ,
\ee
with $c_{n,0,0}=h(\frac{n}{2})$. In \cite{Beisert:2003jj} it was proven that this function is a superconformal invariant and that it has the appropriate eigenvalues when acting on the $\hat{\Cm}_{0(\frac{q+1}{2},\frac{q-1}{2})}$ modules, namely
\be 
H_{12}\hat{\Cm}_{0(\frac{q+1}{2},\frac{q-1}{2})}=2h(q+1)\hat{\Cm}_{0(\frac{q+1}{2},\frac{q-1}{2})}\,.
\ee

\subsection{\textbf{$\Vm \times \Hm \lra \Hm \times {\Vm}$}}

General states in $\Vm \times \Hm$ and $\Hm \times {\Vm}$  can be written as
\bea
|s_1,...,s_n;A\rangle_{\text{\tiny{$\Vm\times\Hm$}}} & = & A^\dag_{s_1 A_1}... A^\dag_{s_n A_n}|0\rangle \otimes |\db\rangle\, ,
\\
|s_1,...,s_n;A\rangle_{\text{\tiny{$\Hm\times\check{\Vm}$}}} & = & A^\dag_{s_1 A_1}... A^\dag_{s_n A_n}|\db\rangle \otimes |\check{0}\rangle\, ,
\eea
where $|\db\rangle = \db^\dag |0\rangle$.
We claim that the action of $H_{12}$ is given by \footnote{To simplify the notation we will omit the delta functions $\delta_{C_1,0} \delta_{C_2,0}$.}
\bea
\nn
H_{12}|s_1,...,s_n;A\rangle_{\text{\tiny{$\Vm\times\Hm$}}} & = & \sum_{s'_1,...,s'_n}c_{n+1,n_{12},n_{21}}|s_1,...,s_n;A\rangle_{\text{\tiny{$\Vm\times\Hm$}}} 
\\
& & +\kappa \sum_{s'_1,...,s'_n}c_{n+1,n_{12},n_{21}+1}|s_1,...,s_n;A\rangle_{\text{\tiny{$\Hm\times\check{\Vm}$}}} 
\eea
and
\bea
\nn
H_{12}|s_1,...,s_n;A\rangle_{\text{\tiny{$\Hm\times\check{\Vm}$}}} & = & \kappa^2 \sum_{s'_1,...,s'_n}c_{n+1,n_{12},n_{21}}|s_1,...,s_n;A\rangle_{\text{\tiny{$\Hm\times\check{\Vm}$}}}
\\
& & -\kappa \sum_{s'_1,...,s'_n}c_{n+1,n_{12},n_{21}+1}|s_1,...,s_n;A\rangle_{\text{\tiny{$\Vm\times\Hm$}}}
\eea
Invariance under the superconformal group is guaranteed by the same arguments given in \cite{Beisert:2003jj}.
The only thing we need to check is that this expression correctly reproduces the $2 \times 2$ matrix given in (\ref{FullHonVxH}). This can be easily done with an algebra software like Mathematica.

Let us work out an example. Consider the action of the Hamiltonian on $\l_{\Im}Q_{\Jm}$ (Lorentz and $SU(2)_L$ indices  are open and go along for the ride).
First, we need to write the state in a ``canonical order" to make sure all our signs are correct,
\be
\l_{\Im}Q_{\Jm}  = \ab^\dag_{(1)}\cb_{(1)\Im}^\dag |0\rangle \otimes \cb_{(2)\Jm}^\dag |\db \rangle = \ab^\dag_{(1)} \cb_{(1)\Im}^\dag \cb_{(2)\Jm}^\dag 
|0\rangle \otimes |\db \rangle\, ,
\ee
\be
Q_{\Im}\lh_{\Jm}  =  \cb_{(1)\Im}^\dag |\db \rangle \otimes  \ab^\dag_{(2)}\cb_{(2)\Jm}^\dag |\check{0}\rangle =- \cb_{(1)\Im}^\dag  \ab^\dag_{(2)} \cb_{(2)\Jm}^\dag 
|\db \rangle \otimes |\check{0}\rangle\, .
\ee
For $\l_{1}Q_{1}$, the action of the Hamiltonian is
\bea
\nn
H_{12}\l_{1}Q_{1} & = & c_{4,0,0}\ab^\dag_{(1)}\cb_{(1)1}^\dag \cb_{(2)1}^\dag |0\rangle \otimes |\db \rangle +c_{4,1,1}\ab^\dag_{(1)}\cb_{(2)1}^\dag \cb_{(1)1}^\dag |0\rangle \otimes  |\db \rangle
\\
\nn
& & +\kappa \left( c_{4,1,1}\ab^\dag_{(2)}\cb_{(1)1}^\dag \cb_{(2)1}^\dag |\db \rangle \otimes |\check{0}\rangle +c_{4,2,2}\ab^\dag_{(2)}\cb_{(2)1}^\dag \cb_{(1)1}^\dag|\db \rangle \otimes |\check{0}\rangle \right)
\\
& = & \l_{1}Q_{1} -\kappa Q_{1}\lh_{1} \, ,
\eea
while  for $\l_{1}Q_{2}$,
\bea
\nn
H_{12}\l_{1}Q_{2} & = & c_{4,0,0}\ab^\dag_{(1)}\cb_{(1)1}^\dag \cb_{(2)2}^\dag |0\rangle \otimes |\db \rangle +c_{4,1,1}\ab^\dag_{(1)}\cb_{(2)1}^\dag \cb_{(1)1}^\dag |0\rangle \otimes  |\db \rangle
+c_{4,1,1}\ab^\dag_{(2)}\cb_{(1)1}^\dag \cb_{(1)1}^\dag |0\rangle \otimes  |\db \rangle
\\
\nn
& &+\kappa c_{4,1,1}\left(\ab^\dag_{(1)}\cb_{(2)1}^\dag \cb_{(2)2}^\dag |\db \rangle \otimes |\check{0}\rangle   
+\ab^\dag_{(2)}\cb_{(1)1}^\dag \cb_{(2)2}^\dag |\db \rangle \otimes |\check{0}\rangle\right) 
\\
\nn
& & +\kappa c_{4,2,2}\ab^\dag_{(2)}\cb_{(2)1}^\dag \cb_{(1)2}^\dag |\db \rangle \otimes |\check{0}\rangle 
\\
& = & \frac{3}{2}\l_{1}Q_{2} -\frac{1}{2}\l_{2}Q_{1} +\frac{1}{2}\f \p-\frac{\kappa}{2}\left( Q_{1}\lh_2 + Q_2 \lh_1 -\p \fh \right)\,.
\eea
Similar calculations can be done for $\l_{2}Q_{1}$ and $\l_{2}Q_{2}$. The final result is
\bea
\nn
H_{12}\l_{\Im} Q_{\Jm} & = & \frac{3}{2}\l_{\Im} Q_{\Jm}-\frac{1}{2}\l_{\Jm}Q_{\Im} -\frac{1}{2}\epsilon_{\Im \Jm}\phi \pt
\\
& & -\frac{\kappa}{2} Q_{\Im} \lh_{\Jm}-\frac{\kappa}{2}Q_{\Jm} \lh_{\Im} -\frac{\kappa}{2}\epsilon_{\Im \Jm}\pt \fh \, ,
\eea
which is consistent with the explicit Feynman diagram calculations of Appendix \ref{twoA}.

\subsection{$\Hm \times \Hm \leftrightarrow \Vm \times \bar{\Vm} \leftrightarrow \bar{\Vm} \times \Vm$}

For these multiplets we have the following states
\bea
|s_1,...,s_n;A\rangle_{\text{\tiny{$\Hm\times\Hm$}}} & = & A^\dag_{s_1 A_1}... A^\dag_{s_n A_n}|\db\rangle \otimes |\tilde{\db}\rangle\, ,
\\
|s_1,...,s_n;A\rangle_{\text{\tiny{$\Vm\times \bar{\Vm}$}}} & = & A^\dag_{s_1 A_1}... A^\dag_{s_n A_n}|\db \tilde{\db}\rangle \otimes |0\rangle\, ,
\\
|s_1,...,s_n;A\rangle_{\text{\tiny{$\bar{\Vm}\times \Vm$}}} & = & A^\dag_{s_1 A_1}... A^\dag_{s_n A_n}|0\rangle \otimes |\db \tilde{\db}\rangle\, .
\eea
Let us consider the $SU(2)_L$ triplet and singlet cases separately. We have found:

\subsubsection{$SU(2)_L$ singlet}

\bea
\nn
H_{12}|s_1,...,s_n;A\rangle_{\text{\tiny{$\Hm\times\Hm$}}} & = &  \sum_{s'_1,...,s'_n}\left( \kappa^2 c_{n,n_{12},n_{21}}-2c_{n+2,n_{12}+2,n_{21}}\right)
|s_1,...,s_n;A\rangle_{\text{\tiny{$\Hm\times\Hm$}}}
\\
\nn
& & + 2\sum_{s'_1,...,s'_n}c_{n+2,n_{12},n_{21}+1}|s_1,...,s_n;A\rangle_{\text{\tiny{$\Vm\times \bar{\Vm}$}}}
\\
& & + 2\sum_{s'_1,...,s'_n}c_{n+2,n_{12}+1,n_{21}}|s_1,...,s_n;A\rangle_{\text{\tiny{$\bar{\Vm}\times \Vm$}}}\, ,
\eea
\bea
\nn
H_{12}|s_1,...,s_n;A\rangle_{\text{\tiny{$\Vm\times \bar{\Vm}$}}} & = &  \sum_{s'_1,...,s'_n}c_{n+2,n_{12},n_{21}}
|s_1,...,s_n;A\rangle_{\text{\tiny{$\Vm\times \bar{\Vm}$}}}
\\
\nn
& & + \sum_{s'_1,...,s'_n}c_{n+2,n_{12}+2,n_{21}}|s_1,...,s_n;A\rangle_{\text{\tiny{$\bar{\Vm}\times \Vm$}}}
\\
& & + \sum_{s'_1,...,s'_n}c_{n+2,n_{12}+1,n_{21}} |s_1,...,s_n;A\rangle_{\text{\tiny{$\Hm\times\Hm$}}} \, ,
\eea
\bea
\nn
H_{12}|s_1,...,s_n;A\rangle_{\text{\tiny{$\bar{\Vm}\times \Vm$}}} & = &  \sum_{s'_1,...,s'_n}c_{n+2,n_{12},n_{21}}
|s_1,...,s_n;A\rangle_{\text{\tiny{$\bar{\Vm}\times \Vm$}}}
\\
\nn
& & + \sum_{s'_1,...,s'_n}c_{n+2,n_{12},n_{21}+2}|s_1,...,s_n;A\rangle_{\text{\tiny{$\Vm\times \bar{\Vm}$}}}
\\
& & + \sum_{s'_1,...,s'_n}c_{n+2,n_{12},n_{21}+1} |s_1,...,s_n;A\rangle_{\text{\tiny{$\Hm\times\Hm$}}}\, .
\eea

\subsubsection{$SU(2)_L$ triplet}

\be
H_{12}|s_1,...,s_n;A\rangle_{\text{\tiny{$\Hm\times\Hm$}}} =  \kappa^2 \sum_{s'_1,...,s'_n} c_{n,n_{12},n_{21}}
|s_1,...,s_n;A\rangle_{\text{\tiny{$\Hm\times\Hm$}}}\, .
\ee

\section{Discussion}
\label{discusionS}
${\cal N}=2$ superconformal symmetry turns out to be more constraining than naively expected:
it fixes the one-loop Hamiltonian of ${\cal N}=2$ SCQCD completely, and that of the interpolating quiver
theory up to a single parameter. 
Knowledge of the full Hamiltonian
should allow to settle the question of one-loop integrability for the ${\cal N}=2$ SCQCD spin chain.
The question is really whether the {\it full }spin chain is integrable.
One-loop integrable subsectors are easy to identify,
but those are trivially isomorphic to  analogous sectors of ${\cal N}=4$ SYM.
Two notable examples of  one-loop  integrable subsectors are the $SU(2|1)$ sector spanned by the letters $\{ \phi, \lambda_{1 \alpha} \}$,
and the $SU(2,1|2)$ sector spanned by the letters $\{  {\cal D}_{+ \dot \alpha}^k \, \phi\; , {\cal D}_{+ \dot \alpha} ^k\,\lambda_{{\cal I} +}\,  \}$:
the one-loop  dilation operator in these sectors is the same as in ${\cal N}=4$ SYM.

Experimental tests of integrability will involve looking for degenerate ``parity pairs'' in the spectrum, as in \cite{Beisert:2003tq, Beisert:2003jj}.
The ultimate proof of  one-loop integrability would be to find an algebraic Bethe ansatz. In ${\cal N}=4$ SYM,
the universal R-matrix of the $SU(1,1)$ subsector uplifts to the $PSU(2,2|4)$ invariant R-matrix
of the full theory \cite{Beisert:2003yb}. In our case, the search for  a candidate R-matrix should start in the $SU(1,1) \times SU(1|1) \times SU(1|1)$
subsector. Work is in progress along these lines.

Another very interesting model that can be studied by our methods  
 is  ${\cal N}=1$ SQCD at the upper edge of the conformal window ($N_f \sim 3 N_c$). This theory
has a large $N$ Banks-Zaks fixed point and can  be studied in perturbation theory.
Its planar one-loop Hamiltonian 
in the scalar sector has been recently  evaluated in \cite{Poland:2011kg}, and shown to coincide with the  Ising model
in transverse magnetic field, which is of course integrable. This however may be a coincidence due to the simplicity
of the scalar sector and it is important to look at the structure of the full theory.
We have identified
a closed $SU(1,1) \times SU(1|1)$ subsector  from which the full spin-chain Hamiltonian
of ${\cal N}=1$ SQCD can be uplifted. It will be interesting to see whether ${\cal N}=1$
superconformal symmetry is in fact fixing the answer uniquely, and whether integrability
extends to   the full Hamiltonian.

Irrespective of integrability, the interpolating quiver theory and its string dual are 
a rich theoretical playground. They have  been explored  from a variety of viewpoints~\cite{Gadde:2009dj,   Gadde:2010zi,   Pomoni:2011, Gadde:2010ku, Rey:2010ry}.
While integrability is broken away from the orbifold point, one retains remarkable analytic control,
and our results are another indication of the intrinsic simplicity of this model.

\section*{Acknowledgements}
It is a pleasure to thank  Niklas Beisert, Carlo Meneghelli, Vladimir Mitev, Jan Plefka, Christoph Sieg, Matthias Staudacher and George Sterman for useful discussions and correspondence.
E.P. wishes to thank the IHES for its warm hospitality as this work was in progress.
The  work of P.L. and L.R. was supported in part by DOE grant DEFG-0292-ER40697 and by NSF grant PHY-0653351-001.
Any opinions, findings, and conclusions or recommendations expressed in this material are those of the authors and do not necessarily reflect the views of the National Science Foundation.
The work of E.P. is supported in part by the Humboldt Foundation.

\appendix

\section{\label{rep-theory} $\Nm=2$ Superconformal Multiplets}
\label{multipletsA}

Detailed studies of  the possible shortening 
conditions for the $\NN=2$ superconformal algebra were performed  in \cite{Dobrev:1985qv,Dobrev:1985qz,Dolan:2002zh}. In this appendix we summarize their findings in Table \ref{shortening},
following the conventions of \cite{Dolan:2002zh}. 
\begin{table}[h]
{\small
\begin{centering}
\begin{tabular}{|c|l|l|l|l|}
\hline 
\multicolumn{4}{|c|}{Shortening Conditions} & Multiplet\tabularnewline
\hline
\hline 
$\BB_{1}$  & $\QQ_{\alpha}^{1}|R,r\rangle^{h.w.}=0$  & $j=0$ & $\Delta=2R+r$  & $\BB_{R,r(0,\bar{j})}$\tabularnewline
\hline 
$\bar{\BB}_{2}$  & $\tilde{\QQ}_{2 \dot{\alpha}}|R,r\rangle^{h.w.}=0$  & $\bar j=0$ & $\Delta=2R-r$  & $\bar{\BB}_{R,r(j,0)}$\tabularnewline
\hline 
$\EE$  & $\BB_{1}\cap\BB_{2}$  & $R=0$  & $\Delta=r$  & $\EE_{r(0,\bar{j})}$\tabularnewline
\hline 
$\bar \EE$  & $\bar \BB_{1}\cap \bar \BB_{2}$  & $R=0$  & $\Delta=-r$  & $\bar \EE_{r(j,0)}$\tabularnewline
\hline 
$\hat{\BB}$  & $\BB_{1}\cap\bar{B}_{2}$  & $r=0$, $j,\bar{j}=0$  & $\Delta=2R$  & $\hat{\BB}_{R}$\tabularnewline
\hline
\hline 
$\CC_{1}$  & $\e^{\alpha\beta}\QQ_{\beta}^{1}|R,r\rangle_{\alpha}^{h.w.}=0$  &  & $\Delta=2+2j+2R+r$  & $\CC_{R,r(j,\bar{j})}$\tabularnewline
 & $(\QQ^{1})^{2}|R,r\rangle^{h.w.}=0$ for $j=0$  &  & $\Delta=2+2R+r$  & $\CC_{R,r(0,\bar{j})}$\tabularnewline
\hline 
$\bar \CC_{2}$  & $\e^{\dot\alpha\dot\beta}\tilde\QQ_{2\dot\beta}|R,r\rangle_{\dot\alpha}^{h.w.}=0$  &  & $\Delta=2+2\bar j+2R-r$  & $\bar\CC_{R,r(j,\bar{j})}$\tabularnewline
 & $(\tilde\QQ_{2})^{2}|R,r\rangle^{h.w.}=0$ for $\bar j=0$  &  & $\Delta=2+2R-r$  & $\bar\CC_{R,r(j,0)}$\tabularnewline
\hline 
$\mathcal{F}$  & $\CC_{1}\cap\CC_{2}$  & $R=0$  & $\Delta=2+2j+r$  & $\CC_{0,r(j,\bar{j})}$\tabularnewline
\hline 
$\bar{\mathcal{F}}$  & $\bar\CC_{1}\cap\bar\CC_{2}$  & $R=0$  & $\Delta=2+2\bar j-r$  & $\bar\CC_{0,r(j,\bar{j})}$\tabularnewline
\hline 
$\hat{\CC}$  & $\CC_{1}\cap\bar{\CC}_{2}$  & $r=\bar{j}-j$  & $\Delta=2+2R+j+\bar{j}$  & $\hat{\CC}_{R(j,\bar{j})}$\tabularnewline
\hline 
$\hat{\mathcal{F}}$  & $\CC_{1}\cap\CC_{2}\cap\bar{\CC}_{1}\cap\bar{\CC}_{2}$  & $R=0, r=\bar{j}-j$ & $\Delta=2+j+\bar{j}$  & $\hat{\CC}_{0(j,\bar{j})}$\tabularnewline
\hline
\hline 
$\DD$  & $\BB_{1}\cap\bar{\CC_{2}}$  & $r=\bar{j}+1$  & $\Delta=1+2R+\bar{j}$  & $\DD_{R(0,\bar{j})}$\tabularnewline
\hline 
$\bar\DD$  & $\bar\BB_{2}\cap{\CC_{1}}$  & $-r=j+1$  & $\Delta=1+2R+j$  & $\bar\DD_{R(j,0)}$\tabularnewline
\hline 
$\mathcal{G}$  & $\EE\cap\bar{\CC_{2}}$  & $r=\bar{j}+1,R=0$  & $\Delta=r=1+\bar{j}$  & $\DD_{0(0,\bar{j})}$\tabularnewline
\hline
$\bar{\mathcal{G}}$  & $\bar\EE\cap{\CC_{1}}$  & $-r=j+1,R=0$  & $\Delta=-r=1+j$  & $\bar\DD_{0(j,0)}$\tabularnewline
\hline
\end{tabular}
\par\end{centering}
}
\caption{Shortening conditions 
and short multiplets for the  $\NN=2$ superconformal algebra.
\label{shortening}
} 
\end{table}

A generic long multiplet  of the $\NN=2$
superconformal algebra   is denoted by $\AA_{R,r(j,\bar{j})}^{\Delta}$. It is generated by the action of the $8$ Poincar\'e supercharges
$\QQ$ and $\tilde {\QQ}$ on a superconformal primary, which by definition is
 annihilated by all the conformal supercharges $\SS$. When  some combination of
the  $\QQ$'s  also annihilates the primary, the corresponding multiplet
is shorter. 
$|R,r\rangle^{h.w.}_{(j,\bar{j})}$ is the highest weight state with eigenvalues $(R, r, j \bar j)$ under the Cartan
generators of the $SU(2)_{R} \times U(1)_{r}$ R-symmetry  and of the Lorentz group.
The multiplet  built on this state is  denoted as $\mathcal{X}_{R,r(j,\bar{j})}$,
where the letter $\mathcal{X}$ characterizes the shortening condition.
The left column of Table \ref{shortening} labels
the condition. 
A superscript on the label  corresponds to the index $\II =1,2$ of the
supercharge that kills the primary:
for example ${\mathcal B}_1$ refers
to ${\mathcal Q}_\alpha^{\ph{\a}1}$. Similarly a ``bar'' on the label refers to the conjugate condition: for example
$\bar{\BB}_{2}$ corresponds to $\tilde Q_{2 \, \dot \alpha}$ annihilating the state;
this would result in the short anti-chiral multiplet $\bar{\BB}_{R,r(j,0)}$, obeying $\Delta = 2 R -r$.
Note that conjugation reverses the signs of $r$, $j$ and $\bar j$ in the expression of the conformal dimension.

\section{Oscillator Representation}
\label{oscillatorsA}

In this appendix we descibe the oscillator representation of the $\Nm=2$ superconformal algebra $SU(2,2|2)$.
We introduce two sets of bosonic oscillators $(\ab^\a,\ab^\dag_\a)$, $(\bb^{\ad},\bb^\dag_{\ad})$ and one
set of fermionic oscillators
$(\cb^{\Im},\cb^\dag_{\Im})$, where $(\a,\ad)$ are Lorentz indices and $\Im$ is an $SU(2)_R$ index. In addition  we will 
need two more ``auxiliary'' fermionic operators $(\db,\db^{\dag})$ and $(\tilde{\db},\tilde{\db}^{\dag})$.
The non-zero (anti)commutation relations are
\bea 
[\ab^\a,\ab^\dag_\b] & = & \delta^\a_\b\, ,
\\
\lbr \bb^{\ad},\bb^\dag_{\bd} \rbr & = & \delta^{\ad}_{\bd}\, ,
\\
\{ \cb^{\Im},\cb^\dag_{\Jm} \} & = & \delta^{\Im}_{\Jm}\, ,
\\ 
\{ \db,\db^\dag \} & =& \{ \tilde{\db},\tilde{\db}^\dag \} = 1\, .
\eea
In this oscillator representation the generators of $SU(2,2|2)$ read
\bea
\Qm^{\ph{k}\Im}_{\a} &=&  \ab^{\dag}_{\a}\cb^{\Im}\, ,
\\
 \Sm^{\ph{k}\a}_{\Im} & = &\cb^{\dag}_{\Im}\ab^{\a}\, ,
 \\
\tilde{\Qm}_{\ad\Im} & =& \bb^{\dag}_{\ad}\cb_{\Im}^{\dag}\, ,
\\
\tilde{\Sm}^{\ad\Im} & =&  \bb^{\ad}\cb^{\Im}\, ,
\\
\Pm_{\a \bd} & = & \ab^{\dag}_{\a} \bb^{\dag}_{\bd}\, ,
\\
 \Km^{\a \bd} & = & \ab^\a \bb^{\bd}\, ,
\\
\Lm^{\ph{\b}\a}_{\b} & = & \ab^\dag_\b\ab^\a-\frac{1}{2}\delta^\a_\b \ab^\dag_\g\ab^\g\, ,
\\
\dot{\Lm}^{\ph{\bd}\ad}_{\bd} & = & \bb^\dag_{\bd}\bb^{\ad}-\frac{1}{2}\delta^{\ad}_{\bd} \bb^\dag_{\gd}\bb^{\gd}\, ,
\\
\Rm^{\ph{\Jm}\Im}_{\Jm} & = & \cb^\dag_{\Jm}\cb^{\Im}-\frac{1}{2}\delta^{\Im}_{\Jm} \cb^\dag_{\Km}\cb^{\Km}\, ,
\\
r & = &  -\frac{1}{2}\cb^\dag_{\Km}\cb^{\Km}+\frac{1}{2}\db^\dag\db+\frac{1}{2}\tilde{\db}^\dag \tilde{\db}\, ,
\\
D & = & 1 + \frac{1}{2}\ab_{\g}^{\dag}\ab^\g + \frac{1}{2}\bb_{\gd}^{\dag}\bb^{\gd}\, ,
\\
C & = & 1 - \frac{1}{2}\ab_{\g}^{\dag}\ab^\g + \frac{1}{2}\bb_{\gd}^{\dag}\bb^{\gd}-\frac{1}{2}\cb^\dag_{\Km}\cb^{\Km}
-\frac{1}{2}\db^\dag\db-\frac{1}{2}\tilde{\db}^\dag \tilde{\db}\,.
\eea
Here $C$ is a central charge that must kill any physical state. It could be eliminated from the algebra by redefining $r+C\rightarrow r$, but it is useful for implementing the harmonic action so we will keep it.
The quadratic Casimir operator is
\bea
\nn
J^2 & = & \frac{1}{2}D^2 + \frac{1}{2}\Lm_{\a}^{\ph{\a}\b}\Lm_{\b}^{\ph{\b}\a}+ 
\frac{1}{2}\dot{\Lm}_{\ad}^{\ph{\ad}\bd}\dot{\Lm}_{\bd}^{\ph{\bd}\ad}- \frac{1}{2}\Rm_{\Im}^{\ph{\Im}\Jm}\Rm_{\Jm}^{\ph{\Jm}\Im}
\\
& & -\frac{1}{2}[\Qm_{\a}^{\ph{\a}\Im},\Sm_{\Im}^{\ph{\Im}\a}]-\frac{1}{2}[\tilde {\Qm}_{\ad\Im}, \tilde{\Sm}^{\ad\Im} ]
-\frac{1}{2}\{\Pm_{\a\bd},\Km^{\a \bd }\}-\frac{1}{2}(r+C)(r+C) \, .
\eea

\subsection{Vector multiplets $\Vm$ and $\bar{\Vm}$}

We define a vacuum state $|0\rangle$ annihilated by all the lowering operators. Then we identify 
\bea 
\Dm^k \Fm & = & (\ab^{\dag})^{k+2} (\bb^{\dag})^k (\cb^{\dag})^0 |0\rangle\, ,
\\
\Dm^k\lambda & = & (\ab^{\dag})^{k+1} (\bb^{\dag})^k (\cb^{\dag})^1 |0\rangle\, ,
\\
\Dm^k \phi & = & (\ab^{\dag})^{k\ph{+0}} (\bb^{\dag})^k (\cb^{\dag})^2 |0\rangle\, ,
\eea
and
\bea 
\Dm^k \bar{\Fm} & = & (\ab^{\dag})^{k} (\bb^{\dag})^{k+2} (\cb^{\dag})^2 \db^{\dag} \tilde{\db}^{\dag} |0\rangle\, ,
\\
\Dm^k\bar{\lambda} & = & (\ab^{\dag})^{k} (\bb^{\dag})^{k+1} (\cb^{\dag})^1 \db^{\dag} \tilde{\db}^{\dag} |0\rangle\, ,
\\
\Dm^k \bar{\phi} & = & (\ab^{\dag})^{k} (\bb^{\dag})^{k\ph{+0}} (\cb^{\dag})^0\db^{\dag} \tilde{\db}^{\dag}  |0\rangle\,.
\eea
For example,
\be
\lambda_{\Im\a} =  \ab^{\dag}_{\a}\cb_{\Im}^{\dag}|0\rangle
\, ,\quad
\bar{\lambda}_{\Im\ad}  =  \bb^{\dag}_{\ad}\cb_{\Im}^{\dag}\db^{\dag} \tilde{\db}^{\dag}|0\rangle \,.
\ee
It's easy to see that all the quantum numbers match, including the zero central charge constraint.

\subsection{Hypermultiplet $\Hm$}

Similarly, for the hypermultiplet we identify
\bea 
\Dm^k Q & = & (\ab^{\dag})^{k} (\bb^{\dag})^k (\cb^{\dag})^1 \db^{\dag}|0\rangle\, ,
\\
\Dm^k \bar{Q} & = & (\ab^{\dag})^{k} (\bb^{\dag})^k (\cb^{\dag})^1 \tilde{\db}^{\dag}|0\rangle\, ,
\\
\Dm^k \psi & = & (\ab^{\dag})^{k+1} (\bb^{\dag})^k \db^{\dag} |0\rangle\, ,
\\
\Dm^k \tilde{\psi} & = & (\ab^{\dag})^{k+1} (\bb^{\dag})^k  \tilde{\db}^{\dag} |0\rangle\, ,
\\
\Dm^k \bar{\psi} & = & (\ab^{\dag})^{k} (\bb^{\dag})^{k+1} (\cb^{\dag})^2\tilde{\db}^{\dag}  |0\rangle\, ,
\\
\Dm^k \bar{\tilde{\psi}} & = & (\ab^{\dag})^{k} (\bb^{\dag})^{k+1} (\cb^{\dag})^2 \db^{\dag}  |0\rangle\,.
\eea

\subsection{Two-letter Superconformal Primaries}
\label{PrimariesA}

By demanding that they are annihilated by all the conformal supercharges and by the
appropriate combinations of Poincar\'e supercharges, 
we have worked out the expressions for the superconformal primaries of the
  irreducible modules that appear on the right hand side of  the tensor products (\ref{HxH})--(\ref{VxVb}).
  The grassmannOps.m oscillator package by Jeremy Michelson and Matthew Headrick was extremely useful for this task.
 We simply quote the results:

\underline{\textbf{$\Vm \times \Vm$:}}
\bea
\bar{\Em}_{2(0,0)} & = & \phi \phi\, ,
\\
\label{VVprimaryq0}
\bar{\Dm}_{\frac{1}{2}(\frac{1}{2},0)} & = & \lambda_{1+} \phi-\phi\lambda_{1+}\, ,
\\
\nn
\hat{\Cm}_{0(\frac{q+1}{2},\frac{q-1}{2})} & = & \sum_{k=0}^{q-1} \frac{(-1)^k}{k+1}\binom{q-1}{k}\binom{q}{k}
\left( \Dm^{q-k-1}\lambda_{1+} \Dm^k\lambda_{2+}-\Dm^{q-k-1}\lambda_{2+} \Dm^k\lambda_{1+}\right)
\\
\nn
& & + \frac{1}{q+1}\bigg(\sum_{k=0}^{q-1} (-1)^k\binom{q-1}{k}\binom{q+1}{k}
\Dm^{q-k-1}\Fm_{\dot{+}\dot{+}} \Dm^k\phi
\\
\label{VVprimaryq}
& &  + \sum_{k=0}^{q-1} (-1)^k\binom{q-1}{k}\binom{q+1}{k+2}
\Dm^{q-k-1}\phi \Dm^k\Fm_{\dot{+}\dot{+}} \bigg)\,.
\eea
For $\bar{\Vm}\times \bar{\Vm}$ the expressions are identical with $(\phi,\l,\Fm)$ replaced by $(\bar{\phi},\bar{\l},\bar{\Fm})$.
The Casimir operator acting on these modules gives
\bea 
J^2_{12} \bar{\Em}_{2(0,0)} & = & 0\, ,
\\
J^2_{12} \hat{\Cm}_{0(\frac{q+1}{2},\frac{q-1}{2})} & = & (q+1)(q+2)\hat{\Cm}_{0(\frac{q+1}{2},\frac{q-1}{2})}, \quad q \ge -1\,.
\eea

\underline{\textbf{$\Vm \times \Hm$:}}
\bea
\label{VHprimaryqm1}
\bar{\Dm}_{\frac{1}{2}(0,0)} & = & \phi Q_1 \, ,
\\
\nn
\hat{\Cm}_{0(\frac{q+1}{2},\frac{q}{2})} & = & \sum_{k=0}^{q} (-1)^k\binom{q}{k}\binom{q+1}{k}
\left(\Dm^{q-k}\lambda_{2+} \Dm^{k}Q_1 - \Dm^{q-k}\lambda_{1+}\Dm^{k}Q_2\right)
\\
\nn
&  & -\sum_{k=0}^{q} (-1)^{k}\binom{q}{k}\binom{q+1}{k+1} \Dm^{q-k}\phi \Dm^{k}\psi_{+}
\\
\label{VHprimaryq}
&  & +q\sum_{k=0}^{q-1}\frac{(-1)^{k}}{k+1}\binom{q-1}{k}\binom{q+1}{k}  \Dm^{q-k-1}\Fm_{++}\Dm^{k}\bar{\tilde{\psi}}_{\dot{+}}\,.
\eea
As before, for $\bar{\Vm}\times H$ we replace $(\phi,\l,\Fm)$ and $(\psi, \bar{\tilde{\psi}})$ by its conjugates.
The action of the Casimir is
\be
\label{CasimirVxH}
J^2_{12} \hat{\Cm}_{0(\frac{q+1}{2},\frac{q}{2})}  =  (q+\frac{3}{2})(q+\frac{5}{2})\hat{\Cm}_{0(\frac{q+1}{2},\frac{q}{2})}, \quad 
q \ge -1\, .
\ee

\underline{\textbf{$\Hm \times {\Vm}$:}}
\bea
\label{HVprimaryqm1}
\bar{\Dm}_{\frac{1}{2}(0,0)} & = & Q_1\fh\, , 
\\
\nn
\hat{\Cm}_{0(\frac{q+1}{2},\frac{q}{2})} & = & \sum_{k=0}^{q} (-1)^k\binom{q}{k}\binom{q+1}{k+1}
\left(\Dm^{q-k}Q_2 \Dm^{k}\lh_{1+}- \Dm^{q-k}Q_1\Dm^{k}\lh_{2+}\right)
\\
\nn
&  & +\sum_{k=0}^{q} (-1)^{k}\binom{q}{k}\binom{q+1}{k} \Dm^{q-k}\psi_{+} \Dm^{k}\fh
\\
\label{HVprimaryq}
&  & +q\sum_{k=0}^{q-1}\frac{(-1)^{k}}{k+2}\binom{q-1}{k}\binom{q+1}{k+1}  \Dm^{q-k-1}\bar{\tilde{\psi}}_{\dot{+}}\Dm^{k}
\check{\Fm}_{++}\,.
\eea

\underline{\textbf{$\Hm \times \Hm$:}}
\bea
\label{HHprimaryqm1}
\hat{\Bm}_1 & = & Q_1\bar{Q}_1\, ,
\\
\nn
\hat{\Cm}_{0(\frac{q}{2},\frac{q}{2})}  & = & \sum_{k=0}^{q}(-1)^k\binom{q}{k}\binom{q}{k}
\left( \Dm^{q-k}Q_1 \Dm^{k}\bar{Q}_2-\Dm^{q-k}Q_2 \Dm^{k}\bar{Q}_1\right)
\\
\nn
&  & +q\sum_{k=0}^{q-1}\frac{(-1)^{k}}{k+1}\binom{q}{k}\binom{q+1}{k}
\Dm^{q-k}\psi_{+} \Dm^{k}\bar{\psi}_{\dot{+}}
\\
\label{HHprimaryq}
&  &-q\sum_{k=0}^{q-1}\frac{(-1)^{k}}{k+1}\binom{q}{k}\binom{q+1}{k}
\Dm^{q-k}\bar{\tilde{\psi}}_{\dot{+}} \Dm^{k}\tilde{\psi}_{+}\, ,
\eea
with
\be
\label{CasimirVxH}
J^2_{12} \hat{\Cm}_{0(\frac{q}{2},\frac{q}{2})}  =  (q+1)(q+2)\hat{\Cm}_{0(\frac{q}{2},\frac{q}{2})}, \quad 
q \ge -1\,.
\ee

\underline{\textbf{$\Vm \times \bar{\Vm}$:}}
\bea
\nn
\hat{\Cm}_{0(\frac{q}{2},\frac{q}{2})} & = &\sqrt{\frac{2(q+2)}{q+1}}\Bigg( \sum_{k=0}^{q}(-1)^k\binom{q}{k}\binom{q}{k}
 \Dm^{q-k}\phi \Dm^{k}\bar{\phi}
\\
\nn
&  & -q\sum_{k=0}^{q-1}\frac{(-1)^{k}}{k+1}\binom{q}{k}\binom{q+1}{k}
\left( \Dm^{q-k}\lambda_{1+} \Dm^{k}\bar{\lambda}_{2\dot{+}}-\Dm^{q-k}\lambda_{2+} \Dm^{k}\bar{\lambda}_{1\dot{+}}\right)
\\
\label{VVbprimaryq}
&  & +q\sum_{k=0}^{q-2}\frac{(-1)^{k}}{k+2}\binom{q+1}{k+1}\binom{q+2}{k} \Dm^{q-k}\Fm_{++} \Dm^{k}\bar{\Fm}_{\dot{+}\dot{+}}\Bigg)\,.
\eea
For $\bar{\Vm} \times \Vm$ we conjugate all fields.



\section{A Sample Field Theory  Calculation}
\label{sampleA}

In this appendix we work out an example of a Feynman diagram calculation of the one-loop dilation operator.
 We consider  the $H_{12}\l_k\bar{\l}_{n-k} \rightarrow \l_k\bar{\l}_{n-k}$ mixing matrix element. For this we 
require finiteness of the correlation function,
\be 
\int d^4x_i e^{-ik_ix_i} \frac{1}{k!(n-k)!}\langle \Dm^k\l(x) \Dm^{n-k}\bar{\l}(x)\bar{\l}(x_1)\l(x_2) \rangle\,.
\ee
The 1P1 Feynman diagrams are given in the second line of Figure \ref{DiagrHHVV1}. The first is a ``gauge emission'' diagram coming from one of the covariant derivatives acting on the field, the second is a standard gauge loop and the last one is a Yukawa loop that contributes to $\l_k\bar{\l}_{n-k} \rightarrow \bar{\l}_{k}\l_{n-k}$ but not to $\l_k\bar{\l}_{n-k} \rightarrow \l_k\bar{\l}_{n-k}$ so we ignore it. To these contributions we have to add one half of the 
self-energy (Figure \ref{DiagrSelf}) of each external leg to obtain the Hamiltonian of the spin chain (see {\it e.g.} Section 2 of \cite{Minahan:2002ve} and Appendix B of \cite{Beisert:2003jj} for more details). We regularize the 
divergent integrals using a momentum cut-off.
The tree level diagram is
\be 
\frac{(-i)^n}{k!(n-k)!}\frac{k_1^{k+1}}{k_1^2}\frac{k_2^{n-k+1}}{k_2^2}\,,
\ee
where $k_1^{k+1}$ and $k_2^{n-k+1}$ are shorthands for $k^{k+1}_{1\, +\dot{+}}$ and $k^{n-k+1}_{2\, +\dot{+}}$. We will usually suppress the indices and the slash, the powers of $k$ and $n-k$ should help avoid confusion. For example, 
\be 
(-k_2-p)^{n-k} \equiv (-k_{2\, +\dot{+}}-p_{+\dot{+}})^{n-k}\,.
\ee
The contribution to the Hamiltonian is minus the coefficient of $\frac{g_{YM}^2N_c}{8\pi^2} \ln \Lambda$ (taking into account the tree level normalization).

For the gauge loop the standard Feynman rules give (factoring out $ig_{YM}^2N_c$)
\be
\nn
\frac{(-1)}{k_1^2 k_2^2}\int \frac{d^4p}{(2\pi)^4} (i(p-k_1))^k(i(-k_2-p))^{n-k}
\frac{[(p-k_1)\bar{\sigma}^{\mu}k_1]_{\a \bd}[(k_2)\bar{\sigma}^{\nu}(p+k_2)]_{\b \ad}\Delta_{\mu \nu}(p)}{(p-k_1)^2(p+k_2)^2}\,.
\ee
For $(\a,\ad)=(+,\dot{+})$ and $(\b,\bd)=(+,\dot{+})$ in Feynman gauge 
we obtain
\be 
\label{lor}
2\frac{(i)^n}{k_1^2 k_2^2}k_{1\, \g \dot{+}}k_{2\, + \dd}\ve^{\gd \dd}\ve^{\g \dt}\sigma^{\l}_{+ \gd}\sigma^{\rho}_{\dt \dot{+}}
\int \frac{d^4p}{(2\pi)^4}\frac{(p-k_1)^k(-k_2-p)^{n-k}}{p^2(p-k_1)^2(p+k_2)^2}(p-k_1)_{\l}(p+k_2)_\rho\,.
\ee
Let's concentrate on the integral, redefining $p \rightarrow -k_2-p$ we get
\be 
\int \frac{d^4p}{(2\pi)^4}\frac{(-k_1-k_2-p)^k p^{n-k}}{p^2(p+k_2)^2(p+k_1+k_2)^2}(p+k_1+k_2)_{\l}p_\rho\,.
\ee
Introducing Feynman parameters and defining
\bea
l & = & p +k_2x+(k_1+k_2)y\, ,
\\
A & = & (k_1+k_2)y+k_2x-(k_1+k_2)\, ,
\\
B & = & (k_1+k_2)y +k_2x\, ,
\eea
we obtain
\be 
2 \int_0^1dx \int_0^{1-x}dy \int \frac{d^4l}{(2\pi)^4}\frac{(A-l)^k (l-B)^{n-k}}{(l^2-\Delta)^3}(l-A)_{\l}(l-B)_\rho\, ,
\ee
where $\Delta$ are leftovers that do not affect the divergent part. 
Now, from this integral four kinds of Lorentz structure can appear: $g_{\l\rho}$, $g_{\l,+\dot{+}}$, $g_{+\dot{+},\rho}$ and $g_{+\dot{+},+\dot{+}}$. Clearly $g_{+\dot{+},+\dot{+}}\equiv 0$. It turns out 
 that $g_{\l,+\dot{+}}$ and $g_{+\dot{+},\rho}$ give also zero when contracted with $\sigma^{\l}_{+ \gd}$ and $\sigma^{\rho}_{\dt \dot{+}}$ respectively (see eq. (\ref{lor})). The only contribution comes from $g_{\l\rho}$. Then,
\be
\nn
2(-1)^{n-k}\int_{0}^{1} dx\int_{0}^{1-x} dy (-k_1-k_2+k_2x+(k_1+k_2)y)^k  (k_2x+(k_1+k_2)y)^{n-k}\frac{g_{\l\rho}}{4}i \frac{\ln \Lambda}{8\pi^2}\,.
\ee
Denoting the Feynman parameter integral by $I(n,k,k_1,k_2)$, the final result is
\be 
2\frac{(-i)^{n+1}}{k_1^2k_2^2}k_{1\, +\dot{+}}k_{2\, +\dot{+}}(-1)^k I(n,k,k_1,k_2) \frac{\ln \Lambda}{8\pi^2}\,.
\ee
The integral $I(n,k,k_1,k_2)$ can be solved analytically,
\be
I(n,k,k_1,k_2) =\frac{(-1)^k(n-k)!k!}{n+2}\sum_{k^{\pr}=0}^{n}\left(\delta_{k=k^{\pr}}+\delta_{k>k^{\pr}}\frac{n-k+1}{n-k^{\pr}+1}
+\delta_{k<k^{\pr}}\frac{k+1}{k^{\pr}+1}\right)\frac{k_1^{k^{\pr}}}{k^{\pr}!} 
\frac{k_2^{n-k^{\pr}}}{(n-k^{\pr})!}\,.
\ee
The contribution to the Hamiltonian is then
\be 
a_{n,k,k'}=-\frac{1}{n+2}\left(\delta_{k=k^{\pr}}+\delta_{k>k^{\pr}}\frac{n-k+1}{n-k^{\pr}+1}
+\delta_{k<k^{\pr}}\frac{k+1}{k^{\pr}+1}\right) \,.
\ee
The other diagrams can be calculated in a similar way, we list the results for completeness.
The self-energy is
\be 
a_{n,k,k'}=\delta_{k=k'}(h(k+1)+h(n-k+1))\, ,
\ee
while 
the gauge ``emission'' diagram gives
\be 
a_{n,k,k'}=-\frac{\delta_{k\neq k'}}{|k-k'|}+\frac{\delta_{k>k'}}{n-k'+1}+\frac{\delta_{k<k'}}{k'+1}\,.
\ee
The sum of the three contributions gives the result quoted in (\ref{ankllbQQFeyn}).

\section{Two Closed Subsectors and the Magnon S-matrix}
	\label{twoA}
The scattering of magnons in the spin chain of the interpolating SCFT was
studied in \cite{Gadde:2010zi, Gadde:2010ku}. The choice of the $\phi/\check \phi$ spin chain vacuum
breaks the symmetry to $SU(2_{\alpha})\times SU(2_{\hat{I}})\times SU(2_{\dot{\alpha}}|2_{I})$, see \cite{Gadde:2010ku} for a detailed
explanation. The scattering of two magnons is given by a factorized two-body S-matrix
\be
S_{SU(2_{\alpha})\times SU(2_{\hat{I}})\times SU(2_{\dot{\alpha}}|2_{I})}= {S}_{SU(2_{\alpha})\times SU(2_{\hat{I}})}\otimes S_{SU(2_{\dot{\alpha}}|2_{I})}\, .
\ee
The  $S_{SU(2_{\dot{\alpha}}|2_{\Im})}$ S-matrix
 describes the scattering of  magnons in the highest weight of $SU(2_{\alpha})\times SU(2_{\hat{\Im}})$
 and is fixed by symmetry to all loops (up to the overall phase), as a function of the single parameter $\kappa$ \cite{Beisert:2005tm, Gadde:2010ku}.
In this appendix we evaluate the one-loop approximation of $S_{SU(2_{\dot{\alpha}}|2_{\Im})}$ for bifundamental magnons,
using the explicit spin chain Hamiltonian, and find agreement with the algebraic analysis of \cite{Gadde:2010ku}.
For this task it is useful to consider a closed subsector, the {\it right} subsector
\be 
\bigg \{\f,\fh,\pb_{\ad\, \hat{\Im}=\hat{1}},\ptb_{\ad\, \hat{\Im}=\hat{1}}, Q_{\Im\, \hat{\Im}=\hat{1}},\Qb_{\Im\, \hat{\Im}=\hat{1}}\bigg\}\,.
\ee
One can also evaluate the one-loop approximation to the other factor of the two-body S-matrix,  ${S}_{SU(2_{\alpha})\times SU(2_{\hat{\Im}})}$,
which is not fixed by symmetry, by considering the \textit{left} closed subsector  
\be 
\bigg \{\f,\fh,\l_{\Im=1\,\a},\lh_{\Im=1\,\a}, Q_{\Im=1\, \hat{\Im}},\Qb_{\Im=1\, \hat{\Im}}\bigg \}\,.
\ee
We have evaluated the Hamiltonian in both the {\it left} and {\it right} sector by direct Feynman diagrams calculations,
finding perfect agreement with the results of sections \ref{perturbativeS} and \ref{algebraicS}.

Our results for both sectors are as follows:

	{\scriptsize{
\begin{eqnarray*}
 & H_{k,k+1}=\\
 & \bordermatrix{
& \f\l & \l\f & \fh\lh & \lh\fh & \l Q & Q\lh & \Qb \l & \lh \Qb & \l\l & \lh\lh \cr
&&&& \cr
\f\l &  2 & -2 & 0 & 0 & 0 & 0 & 0 & 0 & 0 & 0\cr
\l\f & -2 & 2 & 0 & 0 & 0 & 0 & 0 & 0 & 0 & 0\cr
\fh\lh & 0 & 0  & 2\kappa^{2} & -2\kappa^{2} & 0 & 0 & 0 & 0 & 0 & 0\cr
\lh\fh & 0 & 0  & -2\kappa^{2} & 2\kappa^{2} & 0 & 0 & 0 & 0 & 0 & 0\cr
\l Q & 0 & 0  & 0 & 0 & 2 & -2\kappa & 0 & 0 & 0 & 0\cr
Q\lh & 0 & 0  & 0 & 0 & -2\kappa & 2 & 0 & 0 & 0 & 0\cr
\Qb \l  & 0 & 0 & 0  & 0 & 0 & 0 & 2 & -2\kappa  & 0 & 0 \cr
\lh \Qb & 0 & 0 & 0  & 0 & 0 & 0 & -2\kappa & 2  & 0 & 0 \cr
\l\l & 0 & 0  & 0 & 0 & 0 & 0 & 0 & 0 & 4+2\mathbb{K}_l & 0\cr
\lh\lh & 0 & 0 & 0 & 0 & 0 & 0 & 0 & 0 & 0 & (4+2\mathbb{K}_l )\kappa^{2}} \nonumber
\end{eqnarray*}
}}

{\scriptsize{
\begin{eqnarray*}
 & H_{k,k+1}=\\
 & \bordermatrix{
& \f\ptb & \ptb\fh & \pb\f & \fh\pb & \ptb\pb & \pb\ptb \cr
&&&& \cr
\f\ptb  & \frac{3+\kappa^{2}}{2} & -2\kappa  & 0 & 0 & 0 & 0 \cr
\ptb\fh & -2\kappa & \frac{3\kappa^{2}+1}{2} & 0 & 0 & 0 & 0 \cr
\pb\f   & 0 & 0  & \frac{3+\kappa^{2}}{2} & -2\kappa & 0 & 0 \cr
\fh\pb  & 0 & 0  & -2\kappa & \frac{3\kappa^{2}+1}{2} & 0 & 0 \cr
\p\pt   & 0 & 0  & 0 & 0  & 1+3\kappa^{2}+2\kappa^{2}\mathbb{K}_l & 0\cr
\pt\p   & 0 & 0  & 0 & 0 & 0 &  \kappa^2+3+2\mathbb{K}_l} \nonumber
\end{eqnarray*}
}}
{\scriptsize{
\begin{eqnarray*}
 & \oplus
 & \bordermatrix{
& Q \pb & \ptb \Qb & \pb Q & \Qb \ptb \cr
&&&& \cr
Q \pb      &  \frac{1}{2}+\frac{3}{2}\kappa^{2} & -2\kappa^2 & 0 & 0\cr
\ptb \Qb   &  -2\kappa^2 &  \frac{1}{2}+\frac{3}{2}\kappa^{2}  & 0 & 0\cr
\pb Q      &  0 & 0 &\frac{\kappa^2}{2}+\frac{3}{2}& -2\cr
\Qb \ptb   &  0 & 0 &-2& \frac{\kappa^2}{2}+\frac{3}{2}} \,. \nonumber
\end{eqnarray*}
}}

Here we have chosen the gauge parameter $\xi$ such that the self-energy for $Q_{\Im \hat{\Im}}$ is zero,\footnote{$\xi=-1$ if we write the gauge propagator as \be \Delta(k^2)=\frac{1}{k^2}(g_{\mu\nu}-(1-\xi)\frac{k_\mu k_\nu}{k^2})\,.\ee    
    } with this convention the above matrices can be used in conjunction with the scalar sector result of \cite{Gadde:2010zi}.
	The trace operator in Lorentz space is $\mathbb{K}_l=\varepsilon_{\alpha 
	\beta}\varepsilon^{\gamma \delta}$ (using Wess-Bagger conventions). 
	For example,
	\be 
	H_{12} \lambda_{\alpha 1}\lambda_{\beta 1} = 4\lambda_{\alpha 1}\lambda_{\beta 1} 
	- 2\varepsilon_{\alpha \beta}\lambda^{\g}_{1}\lambda_{\g1}\,.
	\ee

\subsection{S-matrix in the  Right Sector}

We can now solve the two-body scattering problem in the {\it right} sector.

\subsubsection{$\ptb$ $\pb$ and  $\pb$ $\ptb$ scattering}

The index structure of the fields implies that there cannot be any transmission, $\ptb$ must always be to the left of $\pb$, the process is pure reflection.
Our results for the four different combinations of fields and indices are summarized in the following table.
\begin{table}[h]
\begin{center}
\begin{tabular}{|l |c| c| c| c| c| c|}
 \hline
 Incoming & Sector & Scattering Matrix  \\
 \hline
 $\ptb\pb$ & $1_{\ad}\otimes 3_L$  & $S(p_1,p_2,\kappa)$   \\
 $\ptb\pb$ & $3_{\ad}\otimes 3_L$  & -1   \\
 $\pb\ptb$ & $1_{\ad}\otimes 3_L$  & $S(p_1,p_2,1/\kappa)$   \\
 $\pb\ptb$ & $3_{\ad}\otimes 3_L$  & -1   \\
\hline
\end{tabular}
\end{center}
\caption{\label{S1} Components of the S-matrix in the {\it right} sector.}
\end{table}
\newline
where
\be 
S(p_1,p_2,\kappa)=-\frac{1+e^{ip_1+ip_2}-2\kappa e^{ip_1}}{1+e^{ip_1+ip_2}-2\kappa e^{ip_2}}\,.
\ee

\subsubsection{$\pb Q$, $Q\pb$, $\bar{Q}\ptb$ and $\ptb\bar{Q}$ scattering}

These processes are a little bit more interesting because we can have reflection and transmission. 
Taking into account all four combinations we obtain
\begin{table}[h]
\begin{center}
\begin{tabular}{|l |c| c| c| c| c| c|}
 \hline
 Incoming & $T$ and $R$ matrices  \\
 \hline
 $\pb Q$        & $T(p_1,p_2,\kappa),R(p_1,p_2,\kappa)$      \\
 $Q \pb$        & $T(p_1,p_2,1/\kappa),R(p_1,p_2,1/\kappa)$  \\
 $\bar{Q}\ptb$  & $T(p_1,p_2,\kappa),R(p_1,p_2,\kappa)$      \\
 $\ptb\bar{Q}$  & $T(p_1,p_2,1/\kappa),R(p_1,p_2,1/\kappa)$  \\
\hline
\end{tabular}
\end{center}
\caption{\label{S2}
Transmission and reflection coefficients in the {\it right} sector.}
\end{table}
where
\bea 
T(p_1,p_2) & = & -\frac{1-e^{-ip_2+ip_1}}{\kappa e^{-ip_2} + \kappa e^{ip_1} -2}\, ,
\\
R(p_1,p_2) & = &  -\frac{1-\kappa e^{-ip_2} -\kappa e^{ip_1} +e^{-ip_2+ip_1}}{\kappa e^{-ip_2} + \kappa e^{ip_1} -2}\,.
\eea
Comparison of Table \ref{S1} amd Table \ref{S2} with equ.(3.12) of \cite{Gadde:2010ku} shows perfect agreement.

\subsection{S-matrix in the { Left} Sector.}

 Our results for $\l \l$ scattering are summarized in  Table \ref{S3} below.
We could not solve the $\l Q$ scattering problem analytically, but one may straightforwardly find numerical results if needed.
\begin{table}[h]
\begin{center}
\begin{tabular}{|l |c| c| c| c| c| c|}
 \hline
 Incoming & Sector & Scattering Matrix  \\
 \hline
 $\l\l$ & $1_{\a}\otimes 3_R$     & $S(p_1,p_2,\kappa=1)$   \\
 $\l\l$ & $3_{\a}\otimes 3_R$     & -1   \\
\hline
\end{tabular}
\end{center}
\caption{\label{S3} Scattering coefficients in the {\it left } sector.}
\end{table}

\newpage

\bibliographystyle{JHEP}
\bibliography{Completebbl}

\end{document}